\documentstyle[12pt,epsf]{article}

\renewcommand{\theequation}{\thesection.\arabic{equation}}

\newcommand{\be}{\small\begin{equation}}
\newcommand{\ee}{\end{equation}\normalsize\vspace*{-0.1ex}}
\newcommand{\bea}{\small\begin{eqnarray}}

\newcommand{\eea}{\end{eqnarray}\normalsize\vspace*{-0.1ex}}
\newcommand{\bdm}{\small\begin{displaymath}}
\newcommand{\edm}{\end{displaymath}\normalsize\vspace*{-0.1ex}}
\newcommand{\beas}{\small\begin{eqnarray*}}

\newcommand{\eeas}{\end{eqnarray*}\normalsize\vspace*{-0.1ex}}

\newcommand{\n}{\noindent}
\newcommand{\eps}{\epsilon}
\newcommand{\intl}{\int\limits}

\newcommand{\lqcd}{\Lambda_{\rm QCD}}
\newcommand{\alq}{\alpha_s(Q)}
\begin{document}

%%%%% BEGIN  TITLEPAGE

\thispagestyle{empty}
\renewcommand{\thefootnote}{\fnsymbol{footnote}}

\setcounter{page}{0}
\begin{flushright} DESY 95--120\\
UM-TH-95-17\\
hep-ph/9506452\\
June 1995  \end{flushright}

\begin{center}
\vspace*{0.3cm}
{\Large\bf
Power corrections and renormalons\\
in Drell-Yan production}\\
\vspace{1.3cm}
{\sc M.~Beneke\footnote{
Address after Oct, 1, 1995: SLAC,
P.O.~Box 4349, Stanford, CA~94309, U.S.A.
}} \\
\vspace*{0.3cm} {\it Randall Laboratory of Physics\\
University of Michigan\\ Ann Arbor, Michigan 48109, U.S.A.}\\[0.6cm]
and\\[0.6cm]
{\sc V.~M.~Braun\footnote{
Address after Sept, 1, 1995: NORDITA, Blegdamsvej 17, DK-2100
Copenhagen, Denmark
}} \\
\vspace*{0.3cm} {\it DESY\\ Notkestr. 85\\ D--22603
Hamburg, Germany}\\[1.4cm]
{\bf Abstract}\\[0.3cm]
\end{center}

\noindent  The resummed Drell-Yan cross section in the
double-logarithmic approximation
suffers from infrared renormalons. Their presence was interpreted as
an indication for non-perturbative corrections of order $\lqcd/(Q(1-z))$.
We find that, once soft gluon emission is accurately taken into account,
the leading renormalon divergence is cancelled by
higher-order perturbative contributions
in the exponent of the resummed cross section. From this evidence,
`higher twist' corrections to the hard cross section
in Drell-Yan production
should intervene only at order $\lqcd^2/((Q^2 (1-z)^2)$ in
the entire perturbative domain $Q (1-z) > \lqcd$. We compare this
result with hadronic event shape variables, comment
on the potential universality of non-perturbative corrections to resummed
cross sections, and on possible implications for phenomenology.

\begin{center}
\vspace*{0.3cm}
{\it submitted to Nuclear Physics B}
\end{center}

\newpage
\renewcommand{\thefootnote}{\arabic{footnote}}
\setcounter{footnote}{0}

%%%%% END OF TITLEPAGE
%%%%% BEGIN SECTION 1
\section{Introduction}

The factorization theorems of QCD \cite{book}
allow a rigorous
treatment of `hard processes', since the non-perturbative
dynamics can be isolated in a few universal distribution and fragmentation
functions.
The classical case for this approach is the Drell-Yan (DY)
process $h_a+h_b\to\mu^+\mu^- + X$, when a muon pair (or,
alternatively, a massive vector boson) with invariant mass $Q^2$ is
produced in the collision of two hadrons, $h_a$, $h_b$, with
invariant mass $s$. The cross section is then given by\footnote{
For simplicity we quote the Born cross section for production
through an intermediate photon only, and neglect the contribution
from $Z^0$.} \cite{book}

\be
\frac{\mbox{d}\sigma}{\mbox{d} Q^2} = \sigma_0\,W(s,Q^2)
\qquad \sigma_0=\frac{4\pi\alpha_{\rm QED}^2}{9 Q^2 s}
\ee
\be\label{factformula}
W(s,Q^2) = \sum_{i,j} \int\limits_0^1\frac{\mbox{d}x_i}{x_i}\,
\frac{\mbox{d} x_j}{x_j} f_{i/h_a}(x_i,Q^2) f_{j/h_b}(x_j,Q^2)\,
\omega_{ij}(z,\alpha_s(Q)) \, +\,{\cal O}
\!\left(\left(\frac{\lqcd}{Q}
\right)^k\right)\,,
\ee

\n where $f_{i/h_a}(x_i,Q^2)$ ($f_{j/h_b}(x_j,Q^2)$) is the
distribution function for a parton $i$ in $h_a$ (parton $j$
in $h_b$) and $z=Q^2/(x_i x_j s)$ is the ratio of invariant masses
of the produced muon pair and the colliding hard partons. The
`hard cross section' $\omega_{ij}(z,\alpha_s(Q))$ can be expanded
as a power series in the strong coupling $\alpha_s(Q)$. The separation
of the Drell-Yan cross section into distribution functions and a
hard cross section is not unique. We will take the DIS scheme for
the distribution functions, so that for quarks and antiquarks,
which is the only case we will be interested in, the distribution
functions are given by the corresponding deep inelastic scattering
(DIS) cross section. In the following we neglect the sum over
parton species and consider only $i=q$, $j=\bar{q}$.

The DY process is one of the simplest, where one can realize,
experimentally, a short-distance dominated dynamics involving
{\em two} large but possibly disparate scales: When $1-z\ll 1$,
the phase space for gluon
emission is restricted, so that the actual scale of the emission
process is $Q (1-z)$ rather than $Q$. Although soft divergences cancel,
large  finite corrections in the form of `+-distributions'
$\alpha_s(Q)^n [(\ln^{2 n-1}(1-z))/(1-z)]_+$ are left over and
it is necessary to resum them to all orders.
As long as $Q (1-z)>\Lambda_{\rm QCD}$, the relevant
contributions from soft and collinear partons are still
amenable to perturbative analysis. The resummation (to
single-logarithmic accuracy) has been completed in \cite{STE87,CAT89}.

The theoretical discussion is simplified in terms of moments. Introducing
the DY scaling variable $\tau=Q^2/s$, eq.~(\ref{factformula}) can
be written as

\be \label{moments}
W(N,Q^2) \equiv \intl_0^1 d\tau\,\tau^{N-1}\,W(\tau,Q^2)
= f_{q/h_a}(N,Q^2) f_{\bar{q}/h_b}(N,Q^2)\,
\omega_{q\bar{q}}(N,\alpha_s(Q))\,,
\ee

\n where $\omega_{q\bar{q}}(N,\alpha_s(Q))$ ($f_{q/h_a}(N,Q^2)$)
denotes moments of the hard cross section (distribution function)
with respect to $z$ ($x_i$). For large $N$, the moments probe
small $1-z$ with the correspondence rule $[(\ln^n(1-z))/(1-z)]_+
\longleftrightarrow \ln^{n+1} N$. The large higher order corrections
$\alpha_s(Q)^n\ln^k N$ ($k\leq 2 n$) can be resummed systematically
by \cite{STE87,CAT89}

\bea \label{sumform}
\omega_{q\bar{q}}(N,\alpha_s(Q)) &=& H(\alpha_s(Q))\,\exp\Bigg[
\intl_0^1 d z\,\frac{z^{N-1}-1}{1-z}\Bigg\{2 \intl_{Q^2 (1-z)}^{
Q^2 (1-z)^2} \frac{d k_t^2}{k_t^2}\,A(\alpha_s(k_t)) \nonumber\\
&&\hspace*{3cm} + \,B(\alpha_s(\sqrt{1-z} Q))\Bigg\}\Bigg] +
{\cal O}\!\left(\frac{\ln N}{N}\right)\,,
\eea

\n provided the functions $A(\alpha_s)$ and $B(\alpha_s)$ (to
be detailed later) are known to some appropriate order.

In this paper we address the question of possible
non-perturbative contributions to the
resummed cross section in eq.~(\ref{sumform}), suppressed by powers of
the large momentum $Q$ (`power
corrections'). We are motivated by the observation
that the integrals
in this equation include the regions of very small momenta $k_t$,
which render the resummation of $\ln N$ sensitive to the infrared (IR)
behaviour of the strong coupling. The resummed expression
depends on the prescription how to deal with the IR Landau pole in the
(perturbative) strong coupling or, in other words, on
the value of an IR cutoff. It has
been found \cite{CON94a,KOR95} that IR effects to the resummed
formulas should intervene already
at the order $\lqcd/(Q(1-z))$, which suggests
that to this accuracy the factorized hard cross section
has to be complemented by (exponentiated) non-perturbative corrections.

If confirmed, this statement is of a considerable theoretical and
phenomenological interest. For phenomenology, it would imply that
non-perturbative corrections decrease so slowly that they are
numerically important for data analysis even at the largest available
energies.
For a theoretician,  the problem is especially interesting
because without guidance of the operator product expansion
there is no immediate classification of power corrections and one must
rely on different methods.

The approach applied in \cite{CON94a,KOR95} and
in this paper utilizes the fact that the insufficiency of perturbation
theory to account for the region of momenta of order
$\lqcd$ is indicated
by the perturbative expansion itself, through the appearence of
another type of large perturbative corrections in {\em large} orders.
These corrections, proportional to $\alpha_s(Q)^n \,n!$ and
known as infrared (IR) renormalons \cite{irr}, cause the divergence
of the perturbative expansion. The numerical value attached to
such a divergent series depends on the summation prescription. Since,
according to current understanding, QCD {\em is} the theory of strong
interactions, this prescription dependence must be cancelled
by all those contributions that elude a perturbative treatment. By
this argument one can determine the order in $\lqcd/Q$ at which
non-perturbative corrections must enter, or, in other words, the
level of theoretical accuracy beyond which perturbative QCD
must fail.
{} For processes that allow an
operator product expansion like in DIS, the renormalon approach
is complementary and less powerful than the operator product
expansion. In the case of resummed cross sections the
consideration of IR renormalons or, equivalently, IR cutoff
dependence in perturbation theory can provide genuinely new
information on the nature of non-perturbative corrections.
The results of \cite{CON94a,KOR95} suggest
that non-perturbative corrections
are expected to affect the resummed DY cross section at the level of
$1/Q$.
Corrections of this size were also detected for hadronic event shape
variables \cite{WEB94,DOK95}.
Since the evolution equations that govern
soft gluon emission are universal, it was proposed that
non-perturbative corrections of order $1/Q$ can be described by a
single parameter common to
all resummed cross sections  \cite{DOK95,AKH95,Ktalk}.

In this paper we reanalyze this problem. Our main result can
be summarized by the statement that the structure of IR renormalons
and power corrections depends crucially on soft gluon emission at
large angles so that the collinear approximation
does not apply. The phase space is then much more complicated
and process-dependent. To obtain information on soft gluon emission
to power-like accuracy, the functions $A$ and $B$ in eq.~(\ref{sumform})
need to be kept to all orders. In particular,
the leading IR renormalon divergence (implying $1/Q$-corrections)
of the resummed DY cross section found in \cite{CON94a,KOR95} appears
as an artefact of
using a finite-order approximation for these functions.
The apparent $1/Q$-ambiguities in the
`standard' factorized formula eq.~(\ref{sumform}) are cancelled
a zero in higher order {\em perturbative} corrections and the
remaining power corrections appear to be of order $N^2\lqcd^2/Q^2$.
This cancellation is in effect similar (though the physics is different)
to the cancellation of the leading $1/m_Q$ ambiguities
between the phase-space and the series of radiative corrections
to the decay widths of heavy particles \cite{BIG94,BBZ,MONT}. Our
results are obtained in a certain approximation to large-order
behaviour. To go beyond this approximation, two-gluon emssion would
have to be analyzed to power-like accuracy. The fact that even
for one-gluon emission the correct IR renormalon structure follows
only from all-order expressions for $A$ and $B$ makes this a vastly
complicated problem.

Since the language connected with large-order behaviour appears
rather formal it is helpful to recall the following correspondence
which reveals more physical insight: For processes
to which the operator product expansion can be applied,
the power corrections deduced from
renormalons can be identified with contributions of
operators of higher dimension (twist).
The above prescription dependence translates into the ambiguity
in the choice of factorization scale below which fluctuations are
called `soft' and therefore should be included into the definition
of the matrix elements of higher-twist operators. For perturbative
contributions this factorization scale acts as an IR cutoff.
Therefore, more generally, quite in analogy to the large perturbative
corrections of type $\alpha_s(Q)^n [(\ln^{2 n-1}(1-z))/(1-z)]_+$
discussed previously,
the large factorials $\alpha_s(Q)^n \,n!$ can also be understood
\cite{BBZ} as remnants from the Kinoshita-Lee-Nauenberg
infrared cancellations,
provided an IR regularization
with an explicit mass scale such as a finite
gluon mass\footnote{The precise correspondence of \cite{BBZ} is true
in a certain approximation, where only diagrams without gluon
self-coupling are considered, so that no difficulties with
gauge-invariance occur. A generalization, although physically
compelling, is unknown, reflecting the present ignorance how to
deal with renormalons on a diagrammatic level, when inclusion of
gluon self-couplings is necessary.} $\lambda$ is used. Instead to factorials
in large orders, one can then restrict attention to cutoff
dependences like $\lambda/Q$, $\lambda^2/Q^2\ln\lambda^2$ in
{\em lowest} order radiative corrections, which is physically
more intuitive. For IR safe quantities, explicit IR divergences
vanish by construction, while the scaling of power corrections can
be determined by setting the IR cutoff to $\lqcd$.
In this language, we find that potential contributions to Drell-Yan
production of order
$\lambda/Q$ from the restriction on the phase space are
exactly cancelled by the contributions of the same order to the
matrix elements.\footnote{More precisely, we find that the matrix
elements can not be expanded in powers of $\lambda/Q$ near the
phase space boundaries, and should be treated exactly.}

The paper is organized as follows: In Sect.~2 we analyze the exponent
in eq.~(\ref{sumform}). We reproduce the results of \cite{CON94a,KOR95}
in the approximation adopted in these works and proceed to show that
low-order truncations of the functions $A$ and $B$ are insufficient
to draw definite conclusions about non-perturbative corrections
to the exponent. We
outline three possible scenarios for the  complete result.
To distinguish the
correct one, an all-order calculation beyond the soft and collinear
approximation is
required. We introduce and justify the corresponding (still approximate)
all-order calculation in Sect.~2.3. Those readers who prefer to think
in terms of IR cutoff dependence rather than large-order behaviour
might ignore subsection 2.3.

In Sect.~3 we
calculate the partonic DY and DIS cross sections with finite gluon
mass $\lambda$ to first order in the strong coupling
and extract the hard scattering function
$\omega_{q\bar{q}}(z,\alpha_s(Q))$.
We find that all potential corrections of order $N\lambda/Q$ cancel, so that
the one-loop perturbative result is protected from IR
contributions to this accuracy. As mentioned previously this calculation
has a dual interpretation in terms of either cutoff dependence or
large-order behaviour. At this stage we can conclude that
at present there is no evidence for nonperturbative corrections of
order $N\lqcd/Q$ to the DY cross section.
To trace the origin of this cancellation, we repeat the calculation
in the soft limit in Sect.~4, but with a cutoff on the
energy and transverse momentum of emitted gluons. This allows us to pin down
the phase space approximation, in which the DLA looses terms of order
$N\lqcd/Q$ and clarifies the relevance of soft emission at large
angles.

In Sect.~5 we follow the approach of \cite{KOR93} to derive the
soft factorization for the DY cross section in terms of Wilson lines.
We calculate the corresponding Wilson line with an arbitrary number of fermion
loop insertions in the gluon line and check that the result satisfies the
correct renormalization group equation (evolution equation).
We find that all relevant anomalous dimensions
(in the $\overline{\rm MS}$  scheme) are entire functions
in the Borel plane and the IR renormalons enter only through the
initial (boundary) condition for the evolution. The apparent
$1/Q$-corrections to
the standard result eq.~(\ref{sumform}) are due to an unfortunate
choice of particular solution of the evolution equation for the
Wilson line which in turn necessitates a more singular homogeneous
solution than actually required. As a result,
the apparent IR renormalon that indicates
$1/Q$-corrections in the Borel transform of the
exponent in eq.~(\ref{sumform})
is cancelled by a `hidden' zero in the sum of Borel transforms
of the (redefined) anomalous dimensions, which becomes manifest only if
these are calculated to all orders.
We discuss a possibility to reformulate the standard factorization
formulas for DY cross sections in such a way that the
cancellation of leading renormalons is explicit.

In Sect.~6 we summarize, compare our results with those for
thrust averages in $e^+ e^-$ annihilation
and comment on potential implications for processes
other than Drell-Yan.
Appendix A contains a rederivation of the cusp
anomalous dimension of the Wilson line, and in Appendix B we compare
typical phase-space integrals for the DY cross section and the thrust
distribution.

%%%%%%%%%%%%%%%%%% SECTION 2 %%%%%%%%%%%%%%%%%%%%%%%%%%%%%%%%%%%%

\section{Anatomy of the exponent}
\setcounter{equation}{0}

In this section we discuss the exponent in the resummed hard cross
section. We rewrite eq.~(\ref{sumform}) as

\be\label{omega}
\omega_{q\bar{q}}(N,\alpha_s(Q)) = H(\alpha_s(Q))\, \exp\left[
E(N,\alpha_s(Q))\right] + R(N,\alpha_s(Q))\,,
\ee

\n where $R(N,\alpha_s(Q))$ vanishes as $N\to \infty$  and the
exponent is given by

\be \label{exponent}
E(N,\alpha_s(Q))=\intl_0^1 d z\,\frac{z^{N-1}-1}{1-z}
\Bigg\{2\!\!\intl_{Q^2 (1-z)}^{
Q^2 (1-z)^2} \!\frac{d k_t^2}{k_t^2}\,A(\alpha_s(k_t))+
B(\alpha_s(\sqrt{1-z} Q))+C(\alpha_s((1-z) Q))\Bigg\}\,.
\ee

\n The exponent expanded in $\alpha_s(Q)$ contains terms
$\alq^n\ln^k N$ with $k\leq n+1$. To sum all logarithms with
$k\geq n+1-m$, the expansion coefficients
of $A(\alpha_s)\equiv \sum_{m=0} a_n\alpha_s^{m+1}$ up to order
$m+1$ and of $B$ and $C$ up to order $m$ are required, as
well as the $\beta$-function that controls evolution of the coupling
to order $\alpha_s^{m+1}$. $A$ is
a process-independent function, which can be identified with
the eikonal
(or cusp) anomalous dimension. Up to second order it is given
by \cite{KOD82}

\be
a_0 = \frac{C_F}{\pi} \qquad a_1 = \frac{C_F}{2 \pi^2}
\left[C_A\left(\frac{67}{18}-\frac{\pi^2}{6}\right) -
\frac{5}{9} N_f\right]\,.
\ee

\n $B$ and $C$ are process-specific. $B$ involves the
underlying DIS process, while $C$ comes only from the DY
process. Their first order expressions
are

\be
b_0=-\frac{3}{2}\frac{C_F}{\pi}\,,\qquad c_0=0\,.
\ee

\n Note that compared to the conventional form \cite{CAT89}
quoted in eq.~(\ref{sumform}) we have reintroduced a third function
$C$ with expansion in the coupling normalized at $(1-z) Q$ as it
was originally present in the analysis of \cite{STE87}. With the
coefficients given above, $C$ is irrelevant for summing logarithms
to next-to-leading accuracy ($k\geq n$) and can be dropped for
this purpose \cite{MAG91,CAT91}. The reason for keeping
this function in the present context will become clear in
Sect.~5. The function $C$ can be dispensed of by a redefinition
of $A$ and $B$ \cite{CAT91}. However, the interpretation of $A$ as
universal cusp anomalous dimension is then lost beyond order
$\alpha_s^2$.

We are now interested in the infrared sensitivity of the
exponent. Perturbatively, it is most directly visible through the
presence of the Landau pole of the perturbative running coupling
inside the integration regions in eq.~(\ref{exponent}). The
Landau pole has two (related) consequences. It leads to IR renormalons
in the re-expansion of $E$ in $\alpha_s(Q)$ and renders the
integrals dependent on the treatment of the pole. The ambiguities
that arise in this way quantify our ignorance on the correct
infrared behaviour, which does not show a Landau pole. Since
extracting the ambiguities from the Landau pole essentially
projects on the infrared regions of integrals, the same conclusions
can be obtained from eliminating these regions by an explicit
IR cutoff and by studying the dependence on the cutoff.
In Sect.~2.1 we reproduce the result on power corrections
in \cite{CON94a} obtained in the double-logarithmic approximation.
The general case is developed in
Sect.~2.2. We introduce
and justify an approximation to calculate large-order
behaviour in Sect~2.3.

\subsection{Ambiguities to double-logarithmic accuracy}

The double-logarithmic approximation (DLA) corresponds to
exponentiation of the $\alq\ln^2 N$-term in $\omega_{q\bar{q}}
(N,\alq)$. In eq.~(\ref{exponent}), only $a_0$ has to be
kept.\footnote{We use the term DLA, if only $a_0$ is kept,
irrespective of whether $\alpha_s$ that multiplies $a_0$ is
frozen or running.}
We remove the contribution from gluons with energy less than
$\mu>\lqcd$. Since $Q (1-z)/2$ can be interpreted as the energy of
the emitted gluon (for small $1-z$), we obtain

\bea\label{dlacutoff}
E^{\rm DLA}(N,\alpha_s(Q),\mu)&=&\intl_0^{1-2\mu/Q}
d z\,\frac{z^{N-1}-1}{1-z}\,\,
2 a_0\!\!\intl_{Q^2 (1-z)}^{
Q^2 (1-z)^2} \!\frac{d k_t^2}{k_t^2}\nonumber\\
&=& {\mbox{$\mu$-independent}} + 2 a_0 \alpha_s(Q)\,
(N-1)\frac{\mu}{Q}\ln\frac{\mu^2}{Q^2}
+ {\cal O}\!\left(\frac{\mu}{Q}\right)\,.
\eea

\n Notice the {\em linear} dependence on the cutoff and the large
coefficient proportional to $N$. In \cite{CON94a} the ambiguity of
the exponent due to the Landau pole has been evaluated in the
approximation of one-loop
running for the coupling,

\be\label{run}
\alpha_s(k) = \frac{\alq}{1-\beta_0\alq\ln(k^2/Q^2)} = \frac{1}{
(-\beta_0)\ln(k^2/\lqcd^2)}\,,
\ee

\n where $\beta_0=-1/(4\pi) (11-2 N_f/3)$ is the lowest-order
coefficient of the $\beta$-function. Following the approximation
of \cite{CON94a} we keep only the term $a_0\alpha_s(k_t)$ up
to terms that give ambiguities of order $1/Q^2$. The $k_t$-integral
produces a cut starting at $z=1-\lqcd/Q$. Evaluating the $z$-integral
above or below the cut, we find the difference (neglecting terms
of order $(\lqcd/Q)^2$)

\bea\label{dlalandau}
\delta E^{\rm DLA}(N,\alpha_s(Q)) &=&
\frac{1}{2\pi i} \intl_0^1 d z\frac{z^{N-1}-1}{1-z}\, 2 a_0
\left(-\frac{1}{\beta_0}\right)\,\mbox{disc}\left[\ln\ln
(1-z)^2 \frac{Q^2}{\lqcd^2}\right]\nonumber\\
&=& 2 a_0 (N-1) \left(-\frac{1}{\beta_0}\right)\frac{\lqcd}{Q}
+{\cal O}\!\left(\left(\frac{N\lqcd}{Q}\right)^2\right)\,.
\eea

\n The cutoff dependence of eq.~(\ref{dlacutoff}) and the ambiguity
in eq.~(\ref{dlalandau}) agree, if we choose $\mu\sim\lqcd$, since
$\alq\ln(\lqcd^2/Q^2)=(-1/\beta_0)$. It can be shown \cite{ALV95}
that the ambiguity does not change qualitatively, if the coefficients
$a_1$ and $b_0$ are included as required for the summation of
next-to-leading logarithms $\alq^n\ln^n N$. We therefore conclude that
with the truncated series for $A$ and $B$ sufficient to sum
next-to-leading logarithms, the exponent in the form of
eq.~(\ref{exponent}) is prescription-dependent by terms of
order $N\lqcd/Q$, if the expression for the running coupling is not
expanded and truncated. At this point, however, it is not clear whether
these terms are artefacts of the approximations made or whether they
should be interpreted as an indicator for `higher twist' effects
of this order as suggested in \cite{CON94a,KOR95}. To illustrate this
point, we note that to the
accuracy of summing next-to-leading logarithms one may replace
\cite{CAT89}

\be\label{replace}
z^{N-1}-1 \to -\Theta\left(1-\frac{e^{-\gamma_E}}{N}-z\right)
\ee

\n It is easy to check that after this substitution the
corresponding exponent $E$ has no ambiguities at all,
unless $N>Q/\lqcd$. Such large moments are excluded from consideration,
since there is no perturbative treatment for them.\footnote{It can
be seen that the ambiguities that arise for $N>Q/\lqcd$ in this case
are not
related to IR renormalons (poles in the Borel transform, see below),
but to the divergence of the Borel integral at $\infty$, when
one of the kinematic variables, here $Q/N$, becomes of order
$\lqcd$, see \cite{BBB1}.}

\subsection{General case}

In higher orders (contributing only to sub-dominant logarithms
$\alq^n\ln^k N$, $k<n$), the replacement in eq.~(\ref{replace})
requires a redefinition of the functions $B$ and $C$. This suggests
that for the purpose of identifying power corrections to the exponent,
higher order coefficients of $A$, $B$ and $C$ are relevant, while
they are not for a systematic summation of logarithms up to a
certain accuracy. The discussion of ambiguities including higher
order coefficients is simplified in terms of Borel transforms. We
write the exponent as

\be\label{borelrep}
E(N,\alq) = \left(-\frac{1}{\beta_0}\right)\intl_0^\infty
d u\,e^{-u/(-\beta_0\alq)}\,B[E](N,u)\,,
\ee

\n where for any function

\be\label{def}
f(\alq) = \sum_{n=0}^\infty f_n\alq^{n+1} \quad\Longrightarrow
\quad B[f](u) = \sum_{n=0}^\infty f_n (-\beta_0)^{-n} \frac{u^n}{n!}
\,.
\ee

\n Note the expansion parameter is $\alq$. If $f$ has a term of
order $\alq^0$, it is treated separately. The ambiguities of $E$
due to the pole in the running coupling are now generated by
IR renormalon poles of $B[E]$ in the $u$-integral. (These poles
correspond to factorial divergence of the corresponding
series expansion in $\alq$.) A pole at $u=m$ leads to an ambiguity
of order $(\lqcd/Q)^{2 m}$. The Borel representation
allows us to write $B[E]$ in a factorized form. Let us continue to
assume that $\beta(\alpha_s)=\beta_0\alpha_s^2$, so that eq.~(\ref{run})
is exact. Then

\be\label{BT}
B[\alpha_s(x Q)](x,u) = \sum_{n=0}^\infty \frac{(-1)^n}{n!} \ln x^2
\, u^n = x^{-2 u}\qquad B[\alpha_s(x Q)^n](x,u) = \frac{u^{n-1}}{n!}
x^{-2 u}\,,
\ee

\n so, for any function $f(\alpha_s(x Q))$,

\be\label{homogen}
B[f(\alpha_s(x Q))](x,u) = x^{-2 u}\, B[f(\alq](u)\,.
\ee

\n The $k_t$- and $z$-integrals that appear in $E$ can then be done and
we obtain

\be\label{borelexp}
B[E](N,u) = 2 \Delta_1(N,u) B[A](u) + \Delta_2(N,u) B[B](u)
+ \Delta_3(N,u) B[C](u)\,,
\ee

\n where all dependence on $N$ is factored into the functions

\be\label{deltas}
\Delta_1(N,u) = \frac{1}{u}\left(\frac{1}{2 u}+\Gamma(N)\left[
\frac{\Gamma(-u)}{\Gamma(N-u)}-\frac{\Gamma(-2 u)}{\Gamma(N-2 u)}
\right]\right)\,,
\ee
\be
\Delta_2(N,u) = \frac{1}{u} + \frac{\Gamma(N)\Gamma(-u)}{
\Gamma(N-u)}\,,\qquad \Delta_3(N,u)=\Delta_2(N,2 u)\,,
\ee

\n which depend only on the form of the exponent, but not on the details
of the functions $A$, $B$, $C$. Note that $\Delta_3=\Delta_2-u\Delta_1$,
so that

\be
B[E](N,u) = \Delta_1(N,u) \left\{2 B[A](u)-u B[C](u)\right\} +
\Delta_2(N,u) \left[B[B](u)+B[C](u)\right]\,.
\ee

\n This is the observation that $C$ is in fact redundant \cite{CAT91}
and can be absorbed into a redefinition of $A$ and $B$. If, as in
Sect.~2.1, we keep only the coefficient $a_0$ of $A$, we obtain close
to $u=1/2$,

\be
B[E^{\rm DLA}](N,u) \stackrel{u\rightarrow 1/2}{=} 2 a_0 (N-1)
\frac{2}{1-2 u}\,.
\ee

\n The ambiguity of $E$ derived from eq.~(\ref{borelrep}) (evaluating the
integral above and below the pole) coincides with eq.~(\ref{dlalandau}).
Without truncation of $A$ and $C$, we obtain

\be
B[E](N,u) \stackrel{u\rightarrow 1/2}{=} 2 (N-1)\frac{1}{1-2 u}
\left[2 B[A]\left(\frac{1}{2}\right)-\frac{1}{2}B[C]\left(\frac{1}{2}
\right)\right]\,,
\ee

\n assuming only that $B[A]$ and $B[C]$ are non-singular at $u=1/2$.
Notice that
the term $\Delta_2 B[B]$ and the first term in square brackets
of $\Delta_1$ which originate from the DIS subprocess
do not lead to a singularity at $u=1/2$. The leading pole is at
$u=1$, indicating power corrections of order $N\lqcd^2/Q^2$,
in agreement with the operator product expansion
for deep-inelastic scattering, which
constrains corrections to the leading-twist approximation to
enter at order $1/Q^2$.

Hence, a conclusion regarding the
presence of a pole at $u=1/2$ can not be obtained from finite-order
approximations\footnote{This situation must be distinguished from the
more familiar statement that the overall normalization of renormalon
divergence is not obtained correctly from diagrams with a single
chain of loops as discussed in Sect.~2.3.
In the present case, even for a single chain, the
overall normalization is incorrect with any finite-order approximation
to $A$.} of $A$ etc., since the residue involves the
Borel transforms at the point $u=1/2$. In other words:
The presence of nonperturbative corrections of order $N\lqcd/Q$
can only be discerned, if the anomalous dimension functions in the
exponent are themselves defined to power-accuracy.  Since the
whole point of the resummation formula eq.~(\ref{sumform}) is
to obtain certain large contributions to $\omega_{q\bar{q}}(N,\alq)$
to all orders by finite order calculations of $A$ etc., there is
no compelling motivation to restrict attention to $E$ alone
instead of the complete $\omega_{q\bar{q}}(N,\alq)$ for the
purpose of identifying power corrections, because finite
order approximations to the exponent are insufficient. But since
the exponent captures all leading singular contributions
$\ln^k N$ (but not singular terms $1/N \ln^k N$), one would
still expect the exponent to indicate also the leading
power corrections $(N\lqcd/Q)^k$ (but not $1/N (N\lqcd/Q)^k$).
Regarding the presence of $N\lqcd/Q$-corrections, the following
three scenarios can be envisaged:

\renewcommand{\theenumi}{\Alph{enumi}}
\begin{enumerate}
\item A renormalon ambiguity of order $N\lqcd/Q$, i.e. a pole at
$u=1/2$ is present. This situation
is realized if $2 B[A](u)-u B[C](u)$ is non-zero at $u=1/2$ and
provided no cancellations with the remainder
$R(N,\alpha_s(Q))$ occur, see C below.
Then a nonperturbative correction of order
$N\lqcd/Q$ is required, and it has to exponentiate.
This scenario is suggested by \cite{CON94a,KOR95}.
\item The pole at $u=1/2$ cancels because

\be \label{zero}
2 B[A]\left(\frac{1}{2}\right)-\frac{1}{2}\, B[C]\left(\frac{1}{2}
\right) = 0\,,
\ee

\n and no ambiguity of order $N\lqcd/Q$ remains. Then no nonperturbative
corrections are {\em required} to this accuracy.
Note that the Borel transforms of $A$, $B$ and $C$ could
themselves have poles. Even in this case $B$ can not participate
in the cancellation (if it occurs) at $u=1/2$, because
$\Delta_2(N,1/2)\sim N^{1/2}$, while the residue of the pole
at $u=1/2$ is of order $N$. Note also that after elimination
of the redundant $C$ to arrive at the conventional eq.~(\ref{sumform}), the
previous equation implies that the Borel transform of $A$ as
it appears in eq.~(\ref{sumform}) has a zero at $u=1/2$. We
will verify (in a certain approximation) that this zero is indeed
present.
\item A further possibility is that the exponent does not incorporate
all  leading (in $N$ and $1/Q$) power corrections and
the pole is cancelled with the remainder $R(N,\alq)$ in
eq.~(\ref{omega}). This can happen if the series for the
remainder looks like (for large $n$)

\be
R(N,\alq) = \sum_{n} \alq^{n+1} (-2\beta_0)^n n!\,\frac{1}{N}
\sum_{k=0}^n \frac{1}{k!} \ln^k N^2\,.
\ee

\n To any finite order in $\alq$ the remainder vanishes as
$N\to\infty$, but its Borel transform

\be
B[R](N,u) = \frac{N^{4 u-1}}{1-2 u}
\ee

\n has a pole at $u=1/2$ with residue proportional to $N$.
A cancellation of this type would again imply that no
nonperturbative corrections of order $1/Q$ are required,
but that the soft factorization techniques break down to
power-like accuracy.
\end{enumerate}

To single out the correct scenario and clarify the size of expected
nonperturbative corrections, $\omega_{q\bar{q}}(N,\alq)$ or
at least the functions $A$, $C$ would have to be calculated to
all orders. Obviously, this requires a certain approximation.
 In the following
sections, we calculate a subset of higher order corrections
and argue that this subset provides some generic insight. Before
introducing this subset in the next subsection, we comment on
the approximation of neglecting higher orders
in the $\beta$-function in the derivation of eq.~(\ref{borelexp}).
When the next coefficient, $\beta_1$, is retained a simple factorized
form as in eq.~(\ref{borelexp}) is not immediate. However, provided
the $\beta$-function itself does not have renormalons (as believed
in the $\overline{\rm MS}$-scheme), inclusion of higher orders are
expected only to turn the poles in $u$ into branch point singularities.
The essence of the arguments above remains unaffected.

\subsection{Higher order approximation}

\begin{figure}[t]
   \vspace{0cm}
   \centerline{\epsffile{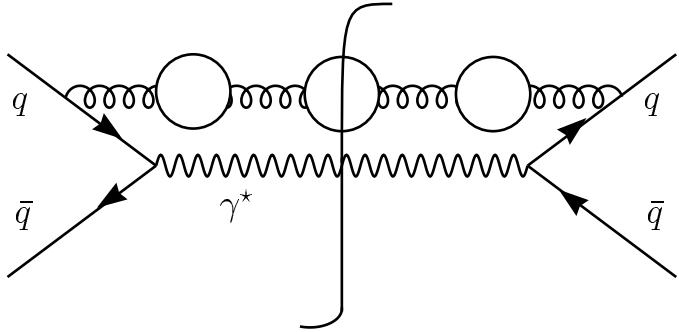}}
   \vspace*{0.3cm}
\caption{\label{dyfig1} $\alpha_s^4$-contribution to the partonic
Drell-Yan cross section. $\gamma^\star$ represents a photon with
invariant mass $Q^2$ that splits into a lepton pair.}
\end{figure}

\phantom{\ref{dyfig1}}

\vspace*{-0.5cm}
In calculating higher-order contributions to the DY hard cross
section, we resort to an approximation that -- despite its largely
heuristic character -- has proven useful in identifying IR renormalons
and power corrections semi-quantitatively in other applications
\cite{MW95,MWI95}. The approximation consists in selecting the subset
of higher-loop diagrams generated by insertion of an arbitrary number
of fermion loops in the gluon line in diagrams contributing
to the lowest order radiative correction, see Fig.~\ref{dyfig1} for
an example. The corresponding series in $\alq$ grows indeed as
$\alq^{n+1} n!$ and its Borel transform has the expected IR renormalon
poles. The approximation (`single-chain approximation') has two immediate
deficiencies which we discuss first. One may keep
in mind that eventually the best justification for the single-chain
approximation derives from its relation to IR cutoff dependence,
which is discussed towards the end of this subsection.

The first observation is that each fermion loop is proportional to
the abelian part of $\beta_0$, i.e. $N_f/(6 \pi)$. The large-order
behaviour from infrared regions is sign-alternating and does not lead
to any ambiguities. Evidently, this happens because with fermion
loops alone, there is no IR Landau pole in the running coupling. The
usual way to deal with this deficiency is to restore the non-abelian
$\beta_0$ by hand.\footnote{This is reflected already in our definition
of the Borel transform, eq.~(\ref{def}), where $\beta_0$ refers to
its non-abelian value.} This is justified by the expectation that the
factorial behaviour is related to evolution of the coupling so that
all other uncalculated diagrams would combine to reproduce this
ad-hoc manipulation, up to an overall normalization, which is of
minor interest for the parametric size of power corrections. The
single-chain approximation then amounts to integrating the lowest
order corrections with the complete (but still approximate) gluon
propagator $\alpha_s(k)/k^2$. This deficiency is not
specific to the DY process, but common
to all previous applications \cite{MW95,MWI95}.

A more specific draw-back of the single-chain approximation is that
within the selected set of diagrams, exponentiation of soft and
collinear logarithms $\ln N$ does not occur. The calculated diagrams
are those with the largest number of factors $N_f$ (the number of
massless fermions) and give at most $\alq^n \ln^{n+1} N$ (This remains
obviously true after restoring the non-abelian $\beta_0$.). The dominant
terms in the large-$N$ limit, $\alq^n\ln^k N$ ($n+1 < k \leq 2 n$,
$n>1$),
come from diagrams with two or more gluon lines (chains). They can be
partially obtained by exponentiation of the single-chain result,
but in any case exponentiation leads outside the set of diagrams
calculated exactly. Thus, if $\omega^{\rm (SC)}_{q\bar{q}}(N,\alq)$
denotes  radiative corrections in the single-chain approximation (SCA),
normalized to the tree-level hard cross section,

\be \ln\left[1+\omega^{\rm (SC)}_{q\bar{q}}(N,\alq)\right]
= \omega^{\rm (SC)}_{q\bar{q}}(N,\alq) + \mbox{terms beyond SCA}\,.
\ee

\n These terms are of order $((\ln N)/N_f))^k$ relative to
$\omega^{\rm (SC)}_{q\bar{q}}(N,\alq)$ so that the large-$N$ limit
(the one we are interested in) does not commute with the
large-$N_f$ limit (the one we calculate).

At this point, we note that the summation of soft and collinear
$\ln N$'s has two aspects: The first is exponentiation as a consequence
of an evolution equation, the second the arguments of the couplings
that appear in the exponent, eq.~(\ref{exponent}). In the present
investigation
of power corrections to the DY process, as well as all previous ones
\cite{CON94a,KOR95}, we are mainly interested
in the {\em exponent} and therefore the second aspect, which is
independent of whether the evolution equation is an equation for
$\ln\left[\omega_{q\bar{q}}(N,\alq)\right]$ or for
$\omega_{q\bar{q}}(N,\alq)$ (in which case exponentiation does not
occur).
Indeed, the single-chain approximation gives a non-trivial
series of higher-order corrections to the anomalous dimension
functions $A$, $B$ and $C$ from subdominant logarithms in diagrams
with fermion loops.
The fact that this series exponentiates beyond
the adopted approximation is secondary as long as one is interested
only in the consequences of integration over the running coupling
in the exponent. One could also say that it is more appropriate to think
of this approximation as an approximation
to $\ln\left[\omega_{q\bar{q}}(N,\alq)\right]$. When interpreted
in this way, corrections are formally of order $1/N_f$ with
no factor of $\ln N$ and the above non-commutativity of the
large-$N$ and large-$N_f$ limit is, at least superficially, absent.

In the context of consequences of integration over the running coupling,
it is more important that the single chain
approximation is consistent with the fact that the appropriate
scale in the coupling to sum large logarithms close to
threshold is given by the transverse momentum of the emitted
parton (gluon) \cite{AMA80}. The real emission correction
to the partonic DY process in this approximation is schematically
given by

\be\label{start}
W^{\rm real}(z,Q^2) \sim \mbox{disc}_{k^2} \intl_0^\infty
\frac{d k^2}{k^2}\int\frac{d |\vec{k}| d ^2 k_t}{(2\pi)^3 2\sqrt{\vec{k}^2
+ k^2}}\,F(z;|\vec{k}|,k_t;k^2)\,\alpha_s(k)\,,
\ee

\n Kinematics dictates that the
discontinuity is integrated up to $(k^2)_{\rm max}\sim
Q^2 (1-z)^2$. Interchanging the order of the $k^2$- and $k_t$-integral,
$k^2$ is integrated up to $k_t^2<Q^2(1-z)^2/4$ (for $z$ close to 1).
Following the argument of \cite{AMA80} the
$k^2$-integral can be evaluated up to subleading logarithms in
$1-z$ by setting $k^2$ to the value at the upper limit of
integration in all terms singular as $k^2\to 0$ and to zero
elsewhere. Therefore

\be \label{amati}
W^{\rm real}(z,Q^2) \sim \int\frac{d^3 \vec{k}}{(2\pi)^3 2
|\vec{k}|}\,F(z;|\vec{k}|,k_t;0)\,\alpha_s(k_t)\,,
\ee

\n up to subleading $\ln(1-z)$'s. The previous line gives simply
the one-loop correction, but with $\alpha_s$ normalized at
$k_t\sim Q(1-z)$.
We see that the single-chain approximation is
fully consistent with \cite{AMA80}: The only approximation is that the
the gluon-propagator that enters the Schwinger-Dyson equation of
\cite{AMA80} is given by $\alpha_s(k)/k^2$ in the single-chain
approximation. In this approximation,
the argument of the coupling is determined
by the subprocesses in which the gluon line splits into the $\bar q q$ pair,
with $k_t^2$ being the largest possible invariant mass of this
pair, which is allowed by the kinematics of soft emission.

Let us emphasize that due to the approximations
made to arrive at eq.~(\ref{amati}), it can not be used
immediately to deduce power-corrections. Also, there is no justification
for integrating eq.~(\ref{start}) with $\alpha_s(k_t)$ instead
of $\alpha_s(k)$, since there is not even a formal limit in which
such a replacement has a diagrammatic interpretation or would
model a complete gluon propagator.

It is useful  to observe that
the Borel transform of radiative corrections in the single-chain
approximation can be calculated as \cite{BBZ,BBB1,BB94b}

\be\label{rel}
B\left[\omega^{\rm (SC)}_{q\bar{q}}\right](N,u) = -\frac{\sin(\pi u)}
{\pi u}\,e^{5 u/3}\intl_0^\infty d \lambda^2 \left(
\frac{\lambda^2}{Q^2}\right)^{-u} \frac{d}{d\lambda^2}\,
\omega^{\rm (1)}_{q\bar{q}}(N,\lambda^2/Q^2)\,,
\ee

\n where $\omega^{\rm (1)}_{q\bar{q}}(N,\lambda^2/Q^2) \alq$
denotes the one-loop correction to the hard cross section,
calculated with a gluon of mass $\lambda$. The factor $5/3$ in the
exponent arises, because we renormalize fermion loops in the
$\overline{\rm MS}$ scheme.

Eq.~(\ref{rel}) gives a transparent interpretation of IR renormalons
in terms of IR cutoff dependence.
By general properties
of Mellin transforms, poles at positive $u$ on the left hand
side of (\ref{rel}) are related to non-analytic terms (in $\lambda^2$)
in the expansion of
$\omega^{\rm (1)}_{q\bar{q}}(N,\lambda^2/Q^2)$ for small
$\lambda^2$. The existence of the pole in question, at $u=1/2$,
is related to the presence of $\sqrt{\lambda^2}$, at $u=1$ to
$\lambda^2\ln \lambda^2$ etc.. Indeed, instead of examining
the large-order behaviour of perturbation theory to find hints for
power corrections, it
appears more direct to explore the IR cutoff dependence in
lowest order. Then, just as an IR divergence $\ln \lambda^2$ in
the partonic DY cross section indicates that the process is
not IR safe but requires the introduction of IR sensitive distribution
functions, the presence of further non-analytic terms indicates
`higher twist' nonperturbative corrections, at least with the
adopted choice of leading-twist distribution functions.

Eq.~(\ref{rel}) is technically convenient, because to
calculate the hard cross section an IR regulator for intermediate
steps is needed even in any case. Taking
finite gluon mass instead of the more conventional dimensional
regularization, the calculation differs from the one-loop
calculation only in that we do not neglect terms that vanish as
$\lambda\to 0$.

%%%%%%%%%%%%%%%%%% SECTION 3 %%%%%%%%%%%%%%%%%%%%%%%%%%%%%%%%%%%%

\section{Hard cross section with IR cutoff}
\setcounter{equation}{0}

In this section we calculate the hard Drell-Yan cross section with finite
IR regulator and investigate its cutoff dependence. The choice of gluon
mass $\lambda$ as regulator is convienent because by eq.~(\ref{rel})
the cutoff dependence can be translated into a statement
on IR renormalons.

The calculation is a textbook problem. Since
$\omega_{q\bar{q}}(z,\alq)$ does not depend on the particular initial
state, one chooses to calculate the DY cross section and structure
functions for quark states. It is convenient to introduce the
dimensionless variable $\xi=\lambda^2/Q^2$. For the partonic DY
and quark distribution function, let us write

\bea
W(\tau,Q^2,\xi) &=& Q_q^2\left[\delta(1-\tau) + W^{\rm (1)}(\tau,\xi)
+ {\cal O}\left(\alq^2\right)\right]\nonumber\\
f_{q/q}(x,Q^2,\xi) &=& \delta(1-x) + f_{q/q}^{\rm (1)}(x,\xi)
+ {\cal O}\left(\alq^2\right)\,,
\eea

\n where $Q_q$ is the electric quark charge in units of electron
charge. In the DIS scheme the distribution function is defined as

\be f_{q/q}(x,Q^2) = \frac{F_2}{x} = -W^\mu_\mu + 12 x^2 \frac{p_\mu
p_\nu}{Q^2} W^{\mu\nu}\,.
\ee

\n $W^{\mu\nu}$ denotes the hadronic tensor for deep inelastic
scattering from quarks with momentum $p$. According to
eq.~(\ref{factformula}), the hard cross section is given by

\bea\label{hard}
\omega_{q\bar{q}}(z,\alq,\xi) &=& Q_q^2 \left[\delta(1-z) +
\omega_{q\bar{q}}^{\rm (1)}(z,\xi)
+ {\cal O}\left(\alq^2\right)\right]\nonumber\\
\omega_{q\bar{q}}^{\rm (1)}(z,\xi) &=& W^{\rm (1)}(z,\xi) - 2
f_{q/q}^{\rm (1)}(z,\xi)\,,
\eea

\n using equality of $f_{q/q}$ and $f_{\bar{q}/\bar{q}}$.

\subsection{Partonic Drell-Yan cross section}

The matrix element for

\be q(p_1) + q(p_2) \longrightarrow g(k) + \gamma^\star(q)
\ee

\n averaged (summed) over initial (final) colours and polarizations
is given by

\be\label{matrixDY}
|{\cal M}|^2_{\rm DY} = 2\frac{C_F g_s^2}{N_c} \left[\frac{2 s
(Q^2+\lambda^2)}{t u}+\frac{t}{u}-\frac{\lambda^2 Q^2}{u^2} +
\frac{u}{t}-\frac{\lambda^2 Q^2}{t^2}\right]\,,
\ee

\n with $s=(p_1+p_2)^2\equiv Q^2/z$, $t=(p_1-k)^2$, $u=(p_2-k)^2$.
Integrating over
the momentum of the photon with invariant mass $Q^2$ gives

\be \label{oneloopint}
W^{\rm (1),real}(z,\xi) = \frac{N_c}{(2\pi)^3} \int\!\frac{d^3 \vec{k}}
{2 \sqrt{\vec{k}^2+\lambda^2}}\,\delta\left((p_1+p_2-k)^2-Q^2\right)
\,|{\cal M}|^2_{\rm DY}\,,
\ee

\n and, finally ($\bar{z}\equiv 1-z$),

\bea\label{realDY}
W^{\rm (1),real}(z,\xi) &=& \frac{C_F\alq}{\pi}\,\Bigg\{\!
\left(\frac{2 (1+\xi) z}{\bar{z}-\xi z}+
\bar{z}-\xi z\right) \ln\frac{\bar{z}-\xi z + \sqrt{
(\bar{z}+\xi z)^2-4\xi z}}{\bar{z}-\xi z - \sqrt{
(\bar{z}+\xi z)^2-4\xi z}}\nonumber\\
&& -\,2 \sqrt{\left(\bar{z}+\xi z\right)^2-4\xi z}
\Bigg\} \,\Theta\!\left(\frac{(1-\sqrt{z})^2}{z}-\xi\right)\,.
\eea

\n The virtual gluon corrections are

\bea\label{virtDY}
W^{\rm (1),virt}(z,\xi) &=& \frac{C_F\alq}{\pi}\,\delta(1-z)\,\Bigg\{
(1+\xi)^2\left[{\rm Li}_2(-\xi)-\frac{1}{2}\ln^2\xi+\ln\xi\ln(1+\xi)
+\frac{\pi^2}{6}\right]\nonumber\\
&& \hspace*{2cm} -\,\frac{3}{2}\ln\xi-\frac{7}{4}-\xi\ln\xi -
\xi\Bigg\}\,.
\eea

\n The double-logarithmic divergences $\ln^2\xi$ from soft and
collinear gluons cancel between real and virtual corrections (in
the sense of distributions). The remaining collinear divergence is
eliminated by subtracting distribution functions.

\subsection{Distribution function}

The matrix element for the DIS process $q(p)+\gamma^\star(q)
\longrightarrow q(\bar{p})+g(k)$
%(fig.~(\ref{dyfig3}))
is given by ($Q^2=-q^2$)

\bea
|{\cal M}|^{2\,\mu}_{{\rm DIS}\,\mu} &=& -4 C_F g_s^2 \left[
\frac{2 u (Q^2-\lambda^2)}{s t}-\frac{t}{s}-\frac{\lambda^2 Q^2}{s^2}
-\frac{s}{t}-\frac{\lambda^2 Q^2}{t^2}\right]\nonumber\\
p^\mu p^\nu\,|{\cal M}|^2_{{\rm DIS}\,\mu\nu} &=& -2C_F g_s^2
\frac{u (t-\lambda^2)^2}{t^2}\,,
\eea

\n with  $s=(p+q)^2$, $t=(p-k)^2$, $u=(\bar{p}-p)^2$. The resulting
real corrections to the quark distribution function read ($\bar{x}\equiv
1-x$)

\bea\label{realDIS}
f^{\rm (1),real}_{q/q}(x,\xi) &=& \frac{C_F\alq}{2\pi}\,\Bigg\{\!
\left[-\frac{1+x^2-2\xi x (1+x-\xi x)}{\bar{x}} - 6\xi x^2
(2-3 \xi x)\right]\ln\frac{\xi x^2}{\bar{x} (1-\xi x)}\nonumber\\
&&\hspace*{1cm}+\,\frac{\bar{x}-\xi x}{\xi x-1} + \frac{
(\bar{x}-\xi x)^2}{2\bar{x}^3}-2 (1-\xi) x\frac{\bar{x}-\xi x}
{\bar{x}^2}+ 3 x\frac{(\bar{x}-\xi x)^2}{\bar{x}^2}\\
&&\hspace*{1cm}-\,12 \xi x^2\frac{\bar{x}-\xi x}{\bar{x}} + 6\xi x
\bar{x} (1-\xi x) - 6\xi^2 x^3\Bigg\}\,\Theta\!\left(\bar{x}-\xi x
\right)\nonumber
\eea

\n The virtual gluon corrections are

\bea
f^{\rm (1),virt}_{q/q}(x,\xi) &=& \frac{C_F\alq}{2\pi}\,
\delta(1-z)\,\Bigg\{
2 (1-\xi)^2\left[{\rm Li}_2(\xi)-\frac{1}{2}\ln^2\xi+\ln\xi\ln(1-\xi)
-\frac{\pi^2}{3}\right]\nonumber\\
&& \hspace*{2cm} -\,3\ln\xi-\frac{7}{2}+ 2 \xi\ln\xi +2 \xi\Bigg\}\,.
\eea

\subsection{Cutoff dependence}

Subtracting the distribution function according to eq.~(\ref{hard}),
we arrive at the well-known IR finite result \cite{ALT78}

\be
\omega^{\rm (1)}_{q\bar{q}}(z,\xi) = \frac{C_F \alq}{2\pi} \Bigg\{
 2 (1+z^2)\left[\frac{\ln(1-z)}{1-z}\right]_+ + \frac{3}{
[1-z]_+} - 4 z-6 +\delta(1-z)\left\{1+\frac{4 \pi^2}{3}\right\}
\Bigg\}
\ee

\n up to terms in $\xi$ that vanish when $\xi\to 0$. We are now
interested in the size of this remainder. Again, it is convenient
to calculate these corrections for the moments rather than the
distribution in $z$. We take moments separately for the DY
cross section and the distribution function. The upper limits in
the $z$- and $x$-integrals are determined by

\be z_{\rm max}=\frac{1}{(1+\sqrt{\xi})^2} = 1-2\sqrt{\xi}+\ldots
\qquad x_{\rm max}=\frac{1}{1+\xi}=1-\xi+\ldots\,.
\ee

\n The linear dependence on $\lambda$ in the upper limit for the DY
process reflects the smaller phase space in this case compared to the
DIS cross section. For $z=x\approx 1$, a gluon with mass

\be \frac{Q^2}{4} (1-z)^2 < \lambda^2 < Q^2 (1-z)
\ee

\n can be emitted in the DIS process but not in DY. For the purpose
of identifying leading power corrections, expansion of the moments
for small $\lambda$ is sufficient. We obtain

\bea\label{momentssmall}
W^{\rm (1)}(N,\xi)&\equiv&\intl_0^1 d z\, z^{N-1}\left[W^{\rm (1),real}
(z,\xi)+W^{\rm (1),virt}(z,\xi)\right]\\
&=& \frac{C_F \alq}{\pi}\Bigg\{\left[\Psi(N)+\Psi(N+2)+2\gamma_E-
\frac{3}{2}\right] \ln\xi + M_{\rm DY} - \left[\frac{N}{2}+1\right]
\xi\ln^2\xi\nonumber\\
&&\hspace*{0.5cm}+\,\left[N^2-N (\Psi(N)+\Psi(N+2)+2\gamma_E) + 3 N
-1\right] \xi \ln\xi + {\cal O}(\xi)\Bigg\}\nonumber\\[0.2cm]
f^{\rm (1)}_{q/q}(N,\xi)&=& \frac{C_F \alq}{2 \pi}
\Bigg\{\left[\Psi(N)+\Psi(N+2)+2\gamma_E-
\frac{3}{2}\right] \ln\xi + M_{\rm DIS}\\
&&\hspace*{0.5cm}+\,\left[N-2 (\Psi(N+1)+\Psi(N+2)+2\gamma_E)+8
-\frac{12}{N+2}\right] \xi\ln\xi + {\cal O}(\xi)\Bigg\}\nonumber\,,
\eea

\n where the $\xi$-independent terms $M_{\rm DY}$ and $M_{\rm DIS}$
are not of interest presently. $\Psi(x)$ is the derivative of the
logarithm of the $\Gamma$-function and $\gamma_E=0.577\ldots$.
For the hard cross section the result reads

\be
\omega^{\rm (1)}_{q\bar{q}}(N,\xi) = \frac{C_F \alq}{\pi}\Bigg\{
M_{\rm DY}-M_{\rm DIS}+\left[\frac{N}{2}+1\right]\xi\ln^2\xi +
\left[N^2+{\cal O}(N\ln N)\right] \xi \ln\xi + {\cal O}(\xi)\Bigg\}
\,.
\ee

\n The terms written out explicitly come entirely from the Drell-Yan
process.

Remarkably, the moments have no contribution linear on the
cutoff $\lambda$ (no $\sqrt{\xi}$-term). Contrary to the what has
been concluded from the double-logarithmic approximation to the
matrix elements, the full hard cross
section for the Drell-Yan process does not indicate `higher-twist'
corrections of order $\lqcd/Q$ from its IR cutoff dependence.
According to Sect.~2.3 this also implies that there is no IR renormalon
located at $u=1/2$ and no ambiguity of order $\lqcd/Q$ in the
hard cross section. With the use of eq.~(\ref{rel}), we find the
leading pole at $u=1$ to be a double pole with residue proportional
to $N$ from the $\xi \ln^2\xi$-term above. For $N>\ln(Q/\lqcd)$ the
single pole with residue proportional to $N^2$ is dominant. From
eq.~(\ref{borelrep}), we obtain the ambiguity (due to the Landau
pole in $\alpha_s$)

\be
\delta\omega_{q\bar{q}}(N,\alq) =
\frac{C_F}{\pi} \left(-\frac{e^{5/3}}{\beta_0}\right)
\left(\frac{N\lqcd}{Q}
\right)^2 \left\{1+\frac{2}{N} \ln\frac{Q}{N\lqcd} + {\cal O}\left(
\frac{1}{N},\frac{N\lqcd}{Q}\right)\right\}\,.
\ee

\n The distribution function $f_{q/q}$ is less singular,
both in $\xi$ and
for large $N$. There is no double pole in the Borel transform and the
expansion runs in $N\xi$ (i.e. $N\lqcd^2/Q^2$) rather than $N^2\xi$ as
for the DY cross section. All results can be translated
into $z$-space by the correspondence $N\leftrightarrow (1-z)^{-1}$
and similarly for the distribution functions in $x$.

For completeness, we give the leading large-$N$ asymptotics of
non-analytic terms in the expansion
of $W^{\rm (1)}(N,\xi)$ for small $\xi$ (This can  most easily be derived
from eq.~(\ref{leadingn}) below.),

\be \label{leadingn2}
W^{\rm (1)}(N,\xi) \stackrel{N\gg 1}{=} \frac{C_F\alq}{\pi}
\left\{2\ln N\ln\xi + \sum_{m=1}^\infty\frac{1}{m [m!]^2}
\,(N^2\xi)^m\ln\xi\right\}\,.
\ee

\n The $\ln\xi$ and $\xi\ln\xi$ terms are in agreement with the
large-$N$ limits of the corresponding terms in eq.~(\ref{momentssmall}).
The double-logarithmic term $\xi\ln^2\xi$ is not seen because its
coefficient is suppressed by $1/N$ compared to the single logarithm.

A similar analysis for the distribution function shows that
the small-$\xi$
expansion runs in $N\xi$ in this case. Thus, the leading IR
corrections for large $N$ for the hard cross section
$\omega_{q\bar{q}}(N,\alq)$ coincide with those inferred from
eq.~(\ref{leadingn2}) except for subtraction of the IR divergent
$\ln\xi$.

It is interesting that
linear terms in $\lambda$
($\sqrt{\xi}$) are absent, although the reduction of the phase space
for real emission is linear in the cutoff. From the phase
space reduction it is evident
that any potential linear term must originate from $z$ close to
one, that is from soft gluons. The cancellation of such terms is
possible, because the distribution $W^{\rm (1),real}(z,\xi)$,
although finite
at the endpoint of the $z$-integration for any finite $\lambda$, is
highly singular in the limit $\lambda\to 0$ in the region where $z$
is close to 1. For the same reason it is not legitimate to expand the
distribution in $\lambda$ before taking the $z$-integral. Expansion
of the integrand would yield a series in $\xi/(1-z)$ which is
singular in the endpoint region. In this sense the
absence of
linear terms is due to a cancellation between the modification
of phase space and the distribution over $z$ close to the endpoint
in $z$.

Let us also add the following important comment: The expansion of
the moments above assumes $N^2\xi < 1$. At the same time we did not
specify that $N$ is large (compared to unity) and the result is
applicable for small as well as large $N$ and therefore indifferent
to the semi-inclusive limit. Since, as a physical IR cutoff,
$\lambda$ should scale with $\lqcd$, our conclusions are valid as
long as $N < Q/\lqcd$. They apply to the limit $N\gg 1$ as long
as one does not enter the domain of very large $N$, where no
perturbative approximation is possible. This is the domain when
the typical transverse momentum of the emitted gluon(s) is of
order or below $\lqcd$. At this point the language of power corrections
looses its meaning, since all such `corrections' are of order unity
and there is no leading term to start with. The absence of
$1/Q$-corrections thus holds over the entire perturbative domain
of $N$.

%%%%%%%%%%%%%%%%%% SECTION 4 %%%%%%%%%%%%%%%%%%%%%%%%%%%%%%%%%%%%

\section{Double-logarithmic versus soft limit}
\setcounter{equation}{0}

In the previous section we concluded that the hard Drell-Yan cross
section in the one-loop approximation does not depend linearly on
the cutoff and that there is no corresponding IR renormalon
indicating a $1/Q$-correction. On the other hand, in Sect.~2.1 we
have seen that such a correction arises from the same one-loop
diagrams in the double-logarithmic approximation (DLA), that is the
phase space region, where the emitted gluon is soft and collinear.
In this section we take a closer look into
this apparent difference.

Rather than a finite gluon mass, we now use a cutoff $\mu$ on
transverse momentum $k_t=\omega \sin\theta$ and energy $\omega$ of
the emitted gluon. We work in the soft approximation $\omega =
Q (1-z)/2 \ll Q$. For comparison
with the DLA, it is sufficient to keep only the first term in
square brackets of the matrix element in eq.~(\ref{matrixDY}). It
can be written as $2 Q^2/k_t^2$. The phase space integral
eq.~(\ref{oneloopint}) is given by

\be
W^{\rm (1), soft}(z,\mu) = \frac{C_F\alq}{\pi}\,\Theta(\omega-
\mu)\intl_{\mu^2}^{\omega^2}\frac{d k_t^2}{k_t^2}\,
\frac{Q}{\sqrt{\omega^2-k_t^2}}\,.
\ee

\n Taking moments,

\be \label{wsoft}
W^{\rm (1), soft}(N,\mu) = 2 \frac{C_F\alq}{\pi} \intl_0^{1-2 \mu/Q}
d z\,z^{N-1}\intl_{\mu^2}^{Q^2 (1-z)^2/4}\frac{d k_t^2}{k_t^2}\,
\frac{1}{\sqrt{(1-z)^2-4 k_t^2/Q^2}}\,.
\ee

\n In the DLA one restricts $k_t\ll \omega$, since only this region
gives rise to $\ln^2 N$. Then the $k_t^2$-term in the square root can
be dropped and we obtain

\be
W^{\rm (1), DLA}(N,\mu) = 2 \frac{C_F\alq}{\pi} \intl_0^{1-2 \mu/Q}
d z\,\frac{z^{N-1}}{1-z}\intl_{\mu^2}^{Q^2 (1-z)^2/4}\frac{d k_t^2}
{k_t^2}\,.
\ee

\n Adding virtual corrections ($z^{N-1}\to z^{N-1}-1$) and subtracting
the distribution functions, one obtains the standard expression,
eq.~(\ref{dlacutoff}). From the previous section it is clear that
a linear term in the cutoff does not arise from the DIS process and
the virtual corrections in DY. We can therefore work with the
previous expression, although the limit $\mu\to 0$ can not be taken.
It is understood that the logarithmic divergences that arise in this
limit would be cancelled by virtual corrections and subtraction of
distribution functions. To find the dependence on the cutoff we take
the derivative

\be \label{firstterm}
\mu^2\frac{d}{d\mu^2} W^{\rm (1),DLA}(N,\mu) =
2 \frac{C_F\alq}{\pi}\intl_0^{1-2 \mu/Q} d z\,\frac{z^{N-1}}{1-z}
\ee

\n and rewrite $z^{N-1}=(1-(1-z))^{N-1}$, then using the binomial theorem.
This gives

\be
\mu^2\frac{d}{d\mu^2} W^{\rm (1),DLA}(N,\mu) \sim
2 \frac{C_F\alq}{\pi}\,2 (N-1)\,\frac{\mu}{Q}\,,
\ee

\n where the symbol `$\sim$' means that we keep only linear terms in $\mu$
(and ignore also $\ln \mu^2$). As anticipated one obtains linear cutoff
dependence from the soft-collinear region $k_t\ll\omega\ll Q$.

Let us now relax the assumption of collinearity and allow $k_t\sim
\omega$. In this case (from eq.~(\ref{wsoft}))

\be
\mu^2\frac{d}{d\mu^2} W^{\rm (1),soft}(N,\mu) =
2 \frac{C_F\alq}{\pi}\intl_0^{1-2 \mu/Q} d z\,\frac{z^{N-1}}
{\sqrt{(1-z)^2-4 \mu^2/Q^2}}\,.
\ee

\n The square root can in fact not be approximated. When expanded

\be
\frac{1}{\sqrt{(1-z)^2-4 \mu^2/Q^2}} = \frac{1}{1-z} +
\frac{2\mu^2}{Q^2 (1-z)^3} + \ldots\,,
\ee

\n all terms are of order one in $\mu$ after integration over $z$.
We can still use
this expansion, provided we resum all terms of order $\mu$. As before, we
first obtain

\bea
\mu^2\frac{d}{d\mu^2} W^{\rm (1),soft}(N,\mu) &=&
2 \frac{C_F\alq}{\pi} \sum_{k=0}^\infty\frac{1}{k!}\frac{\Gamma(1/2)}
{\Gamma(1/2-k)}\left(-\frac{4\mu^2}{Q^2}\right)^k\nonumber\\
&&\,\times\,\sum_{l=0}^{N-1}
\left(\!\begin{array}{c}
N-1\\l\end{array}\!\right) \frac{(-1)^l}{l-2 k}\left(\left(\frac{2\mu}{Q}
\right)^{l-2 k}-1\right)
\,.\eea

\n This expression is invalid for $l=2 k$. However, we do not need this
case, because linear terms in $\mu$ originate only from $l=1$. Thus

\be
\mu^2\frac{d}{d\mu^2} W^{\rm (1),soft}(N,\mu) \sim
2 \frac{C_F\alq}{\pi} \,(N-1)\,\frac{\mu}{Q}
\sum_{k=0}^\infty\frac{(-1)^k}{k!}\frac{\Gamma(1/2)}
{\Gamma(3/2-k)}\,.
\ee

\n The DLA gives precisely the first term in the sum. Recognizing
the coefficients of the series as the Taylor-coefficients of
$(1+x)^{1/2}$, the sum of all terms gives

\be
\mu^2\frac{d}{d\mu^2} W^{\rm (1),soft}(N,\mu) \sim
2 \frac{C_F\alq}{\pi} \,2 (N-1)\,\frac{\mu}{Q}\, (1+(-1))^{1/2}
= 0\,.
\ee

\n All linear dependence on $\mu$ has disappeared. Consistent with the
exact calculation of the hard cross section, power corrections are of
order $N^2 \mu^2/Q^2$. As it turned out, corrections of order $N\mu/Q$
arise also from the regions of phase space when a soft gluon is emitted
with large angle ($k_t\sim\omega$)
and no corrections are present after adding all regions.
Another way to state this conclusion is that corrections of order $1/Q$
arise as an artefact of the collinear approximation, which is valid
to the accuracy of leading logarithms in $N$, but not to power-like
accuracy.
This result has important implications: While soft and collinear
gluon emission has many universal features that allow to resum leading large
logarithms for many processes by a single
universal anomalous dimension, the same universality
does not extend to power corrections $N \lqcd/Q$. Depending on the
particular process considered the corrections
of this type inferred from the soft and collinear region might or
might not be cancelled by those from other regions.

%%%%%%%%%%%%%%%%%% SECTION 5 %%%%%%%%%%%%%%%%%%%%%%%%%%%%%%%%%%%%

\section{Soft factorization in the single-chain approximation}
\setcounter{equation}{0}

The DY cross section calculated
to first order in $\alpha_s$ with an explicit IR regulator does not
contain IR contributions linear in the cutoff. In this section we
translate this result into the language of large-order
behaviour in perturbation theory. This serves as a check that the
absence of contributions of order $1/Q$ is
consistent with perturbative factorization \cite{STE87,CAT89,KOR93},
and elucidates which of the options A, B, C, of Sect.~2.2 is realized.
We have seen that for the cancellation of contributions proportional to
$N/Q$ it is necessary to account exactly for soft gluon emission at
all angles. The large logarithms in $N$ beyond the
DLA also exponentiate \cite{STE87,CAT89,KOR93} and
in fact correct for soft gluon
emission at large angles, but
it is only after resummation of all subdominant logarithms
that the result is adequate to exhibit the cancellation of all
IR contributions of order $1/Q$.
Hence, to clarify the situation the anomalous dimension functions
in the exponent of the resummed
cross section have to be calculated to all orders in perturbation theory,
in agreement with general considerations in Sect.~2.2.

Since we do not expect the DIS subprocess to cause power corrections
of order $1/Q$, we restrict ourselves to the contributions from
the DY process, contained in the functions $A$ and $C$ in
eq.~(\ref{exponent}). Being interested only in the soft gluon region,
we recall that the emission of soft gluons from the incoming quark
and anti-quark can be accounted for by eikonal phases equivalent
to Wilson lines along the classical trajectories of the partons
\cite{KOR93}. This is of particular interest in the present
context, since the Wilson line operators take into account all soft
contributions, including the non-collinear ones. In view of
the previous discussion, we would therefore
not expect any $1/Q$-corrections to emerge in this approach.

\subsection{Wilson lines}

We follow the treatment of Korchemsky and Marchesini \cite{KOR93},
generalizing it to our particular
class of higher order contributions. For more details, we refer the
reader to the second reference of \cite{KOR93}.

The matrix element for emission
of soft partons is given by $\langle n|T U_{\rm DY}(0)|0\rangle$,
where

\be
U_{\rm DY}(x) = P\exp\left(i g_s\intl_{-\infty}^0 d s\,p_2^\mu
A_\mu(p_2 s+x)\right)\, P\exp\left(-i g_s\intl_{-\infty}^0
d s\,p_1^{\mu} A_\mu(p_1 s+x)\right)\,,
\ee

\n is the product of Wilson line operators describing the annihilation
 of an on-shell quark and anti-quark with momenta $p_1$ and
$p_2$ at the space-time point $x$.
Up to corrections that vanish as
$z\to 1$, the partonic Drell-Yan cross section is given by

\be\label{softfactorization}
W(z,\alq) =  H_{\rm DY}(\alq)\, W_{\rm DY}(z,\alq)\,.
\ee

\n $H_{\rm DY}=1+{\cal O}(\alq)$ is a short-distance dominated function,
independent of $z$.
$W_{\rm DY}$ is the square of the
matrix element, summed over all final states:

\be
W_{\rm DY}(z,\alq) = \frac{Q}{2}\intl_{-\infty}^
\infty\frac{d y_0}{2\pi}\,e^{i y_0 Q (1-z)/2}\,
 \langle 0|\bar{T}\,U^\dagger_{\rm DY}(y)
\,T\,U_{\rm DY}(0)|0\rangle
\ee

\n The Fourier transform is taken with respect to the energy of soft
partons and $y=(y_0,\vec{0})$.

The crucial observation is that the Wilson line depends only on
the ratio $(\mu N)/(Q N_0)$ (taking moments of $W_{\rm DY}(z,\alq)$,
where $\mu$ is a
cutoff separating soft and hard emission (the renormalization scale
for the Wilson line) and $N_0$ is a suitable constant.
Hence the $N$-dependence of the Wilson lines can be obtained from their
$\mu$-dependence, which is given by the renormalization group equation

\be\label{evolutionequation}
\left(\mu^2\frac{\partial}{\partial\mu^2} + \beta(\alpha_s)
\frac{\partial}{\partial \alpha_s}\right)\,\ln W_{\rm DY}
\left(\frac{\mu^2 N^2}{Q^2 N_0^2},\alpha_s(\mu)\right) =
\Gamma_{\rm cusp}(\alpha_s)\,\ln\frac{\mu^2 N^2}{Q^2 N_0^2}
+ \Gamma_{\rm DY}(\alpha_s)\,.
\ee

\n Here $\Gamma_{\rm cusp}(\alpha_s)$ is the universal cusp anomalous
dimension of the Wilson line \cite{KOR87}. The general solution to
eq.~(\ref{evolutionequation}) is given
by

\be\label{sol1}
\ln W_{\rm DY}\!
\left(\frac{\mu^2 N^2}{Q^2 N_0^2},\alpha_s(\mu)\right) =
X(\alpha_s(Q N_0/N))\,+\!\!\!\!\!\intl_{Q^2 N_0^2/N^2}^{\mu^2}
\!\!\!\!\frac{d k_t^2}{
k_t^2}\left[\Gamma_{\rm cusp}(\alpha_s(k_t)\,\ln\frac{k_t^2 N^2}{
Q^2 N_0^2} + \Gamma_{\rm DY}(\alpha_s(k_t))\right]\,.
\ee

\n
The integral is a particular solution of the inhomogeneous
eq.~(\ref{evolutionequation}) and $X(\alpha_s(Q N_0/N))$ denotes the
general solution of the corresponding homogeneous equation,
an arbitrary function of the running coupling $\alpha_s(Q N_0/N)$.
Putting $\mu=Q N_0/N$, one identifies $X(\alpha_s(Q N_0/N)) =
\ln W_{\rm DY}(1,\alpha_s(Q N_0/N))$. The inhomogeneous term can be
rewritten (identically) in a more familiar form and we obtain for the partonic
Drell-Yan cross section

\bea\label{form1}
W(N,\alq) &=& H_{\rm DY}(\alq)\,W_{\rm DY}(1,\alpha_s(Q N_0/N))\,
\exp\Bigg(2\intl_0^{1-N_0/N}\frac{d z}{1-z}\nonumber\\
&&\hspace*{0.2cm}\intl_{(1-z)^2 Q^2}^{Q^2}
\frac{d k_t^2}{k_t^2}\,\Gamma_{\rm cusp}(\alpha_s(k_t))
+\,\Gamma_{\rm DY}(\alpha_s((1-z) Q)\Bigg)
+ {\cal O}(\ln N/N)\,.
\eea

\n We have restored the hard part $H_{\rm DY}(\alq)$ and set
$\mu=Q$. This equation differs in form from the standard resummed cross
section eq.~(\ref{omega}) with the exponent eq.~(\ref{exponent}) written
with the replacement eq.~(\ref{replace}) only by the presence of the
{\em initial condition} $W_{\rm DY}(1,\alpha_s(Q N_0/N))$. Its
expansion in $\alpha_s$
produces subdominant logarithms $\alpha^k\ln^{k-1} N$.
To fully conform with the conventional expression for the resummed cross
section, the initial condition can be absorbed into
the following redefinitons of $H_{\rm DY}$ and $\Gamma_{\rm DY}$:

\bea\label{redef}
H_{\rm DY}(\alq) &\longrightarrow& H_{\rm DY}(\alq) W_{\rm DY}
(1,\alq))\nonumber\\
\Gamma_{\rm DY}(\alpha_s) &\longrightarrow& \Gamma_{\rm DY}(\alpha_s)
- \beta(\alpha_s)\frac{d}{d\alpha_s}\,\ln W_{\rm DY}
(1,\alpha_s)
\eea

\n The redefined $\Gamma_{\rm DY}$ starts at order $\alpha_s^2$.
It does not affect resummation of large
logarithms in $N$ to next-to-leading accuracy $\alq^k\ln^k N$.
 The argument of the coupling in $\Gamma_{\rm DY}$ is naturally
$(1-z) Q$ and not  $\sqrt{1-z} Q$ as in the
function $B$ in eq.~(\ref{exponent}). It is for this reason that
we have reintroduced the function $C$ there. If we absorbed $C$ into
a redefintion of $A$ and $B$ as in \cite{CAT91}, $A$ would no
longer coincide with the universal function $\Gamma_{\rm cusp}$
starting from order $\alpha_s^3$.

In the remainder of this section we reproduce the
soft factorization formula (\ref{softfactorization}) to all orders
in the strong coupling, but in the single-chain approximation. We do so
first by taking the large-$N$ limit of the result obtained in Sect.~3,
and then by direct calculation of the Wilson line. We will see that\\

- IR renormalons in the
Wilson line are in exact correspondence with the IR renormalons in the
full cross section in the large-$N$ limit. This excludes the scenario C
in Sect.~2.2 which would have involved the remainder $R$, which is
suppressed by $1/N$ to any finite order in perturbation theory
and not captured by eq.~(\ref{softfactorization}).\\

- the Wilson line satisfies the correct
RG equation, with the anomalous dimensions $\Gamma_{\rm cusp}(\alq)$,
$\Gamma_{\rm DY}(\alq)$ being entire functions in the $\overline{\rm MS}$
scheme. IR renormalons enter exclusively
via the {\em initial condition}
for the evolution of the Wilson line, which by comparison with
the calculation in the single-chain approximation requires non-perturbative
corrections only at the level of $1/Q^2$. This verifies the
absence of $1/Q$-corrections to the DY cross
section within the soft factorization technique.\\

- the standard exponentiated cross section chooses a particular
solution to the evolution equation that requires a redefinition of
both the anomalous dimension functions and
the boundary condition in such a way that both of them contain an
IR renormalon (pretending $1/Q$-ambiguities) that was not present in
the initial formulation and which cancels in the sum.
This cancellation selects scenario B from Sect.~2.2.

\subsection{Large-$N$ limit of the DY cross section}

In this subsection we calculate the cross section $W(z,\alq)$ on the
l.h.s. of eq.~(\ref{softfactorization}) as the large-$N$ limit of the
result obtained in Sect.~3. Instead of calculating the series of higher
orders in $\alpha_s$, it is more convenient to
study the Borel transform, eq.~(\ref{BT}), which can be obtained by
Mellin transformation, eq.~(\ref{rel}),
from the one-loop result with finite gluon mass,
as given in Sect.~3.1. The virtual corrections are $N$-independent
and of no importance for what follows.
For the contribution from real emission we obtain

\bea
B\left[W^{\rm real}\right](N,u) &=& -\frac{C_F}{\pi}
\frac{\sin(\pi u)}
{\pi u}\,e^{5 u/3}\intl_0^\infty d\xi\,\xi^{-u} \,\frac{d}{d\xi}
\intl_0^1 d z\,\Theta\!\left(\frac{(1-\sqrt{z})^2}{z}-\xi\right)\,
\nonumber\\
&& \times \,z^{N-1}\,\frac{2}{\bar{z}-\xi z}
\ln\frac{\bar{z}-\xi z + \sqrt{
(\bar{z}+\xi z)^2-4\xi z}}{\bar{z}-\xi z - \sqrt{
(\bar{z}+\xi z)^2-4\xi z}}\,,
\eea

\n neglecting all terms that do not contribute to the leading
asymptotics
at $N\to \infty$. After repeated substitution of variables, the
integral is

\bea
B\left[W^{\rm real}\right](N,u) &=& \frac{C_F}{\pi} \frac{\sin(\pi u)}
{\pi u}\,e^{5 u/3}\, 2 N \intl_0^1 d x\,\ln\frac{1+\sqrt{x}}{
1-\sqrt{x}}\,(1-x)^{-1-u}\nonumber\\
&& \hspace*{-1.5cm}\times \Bigg[ B(1-2 u,2 N+2 u+1)
{}_2 F_1\left(\frac{1}{2}-u,1-u;N+\frac{3}{2};x\right) \nonumber\\
&&\hspace*{-1cm} + \,
(1-x) B(2-2 u,2 N+2 u)\,{}_2 F_1\left(\frac{3}{2}-u,1-u;N+\frac{3}{2};
x\right)\Bigg]\,,
\eea

\n where $B(x,y)=\Gamma(x)\Gamma(y)/\Gamma(x+y)$ and ${}_2 F_1(
a,b;c;x)$ denotes the hypergeometric function. Because of the arguments
of $B$, the second term in square brackets
contributes only at relative order $1/N$
can be dropped. The first term can be
integrated using

\bea
&&\frac{N+1/2}{u (N+u)}\,\frac{d}{d x}\left[(1-x)^{-u}
{}_2 F_1\left(\frac{1}{2}-u,-u;N+\frac{1}{2};x\right) \right]
\nonumber\\
&&\hspace*{2cm}
= (1-x)^{-1-u}
{}_2 F_1\left(\frac{1}{2}-u,1-u;N+\frac{3}{2};x\right)\,.
\eea

\n The result is

\bea\label{leadingn}
B\left[W^{\rm real}\right](N,u) &=& \frac{C_F}{\pi} \frac{\sin(\pi u)}
{\pi u}\,e^{5 u/3}\,\frac{\Gamma(-u)^2\Gamma(N+u)^2}{\Gamma(N)^2}
\nonumber\\
&\stackrel{N\gg 1}{=}& \frac{C_F}{\pi}\,e^{5 u/3}\,\frac{1}{u^2}
\frac{\Gamma(1-u)}{\Gamma(1+u)}\,N^{2 u}\,.
\eea

\n Close to the poles at integer $u=m$ the result is accurate to
the dominant power of $N$, i.e. $N^{2 m}$, in the residue, neglecting
corrections of order $(\ln N)/N$. The double
pole at $u=0$ is cancelled by the virtual corrections. Poles at
half-intergers do not arise.

A similar analysis for the DIS distribution function shows that the
$N$-dependence of the Borel transform in the large-$N$ limit
is given by $N^u$ in
accordance with the observation in Sect.~3.3 that the small-$\xi$
expansion runs in $N\xi$ in this case.

\subsection{Calculation of the Wilson line}

Next, we obtain the Wilson line $W_{\rm DY}$.
The Wilson line operators need renormalization and we choose
$\overline{\rm MS}$ subtractions. This implies in particular that all
integrals without a scale are set to zero. Since $p_1^2=p_2^2=0$,
all virtual corrections  to $W_{\rm DY}$ vanish and non-zero
contributions arise only from
contraction of gluon fields with different time ordering. The
leading-order expression is then given by

\bea\label{lowestorder}
W^{\rm (1)}_{\rm DY}(z,\alpha_s) &=& C_F g_s^2\,\frac{Q}{2}\,\left(
\frac{\mu^2 e^{\gamma_E}}{4\pi}\right)^\eps \int\frac{d^d k}{(2\pi)^d}\,
2\pi\,\delta(k^2)\,\delta\left(k^0-\frac{Q}{2} (1-z)\right)\,
\frac{2 p_1\cdot p_2}{p_1\cdot k \,p_2\cdot k}\nonumber\\
&=&\frac{C_F\alpha_s}{\pi}\frac{2}{1-z}\left(\frac{Q^2 e^{\gamma_E}}{
4\mu^2} (1-z)^2\right)^{-\eps}\,\frac{\sqrt{\pi}}{(-\eps)
\Gamma(1/2-\eps)}
\eea

\n Note that we have inserted the factor $e^{\gamma_E}/(4\pi)$
in front of $\mu^2$.
With this choice of renormalization scale $\mu$, $\overline{\rm MS}$
corresponds to subtraction of poles in $\eps$ ($d=4-2\eps$).

\begin{figure}[t]
   \vspace{0cm}
   \centerline{\epsffile{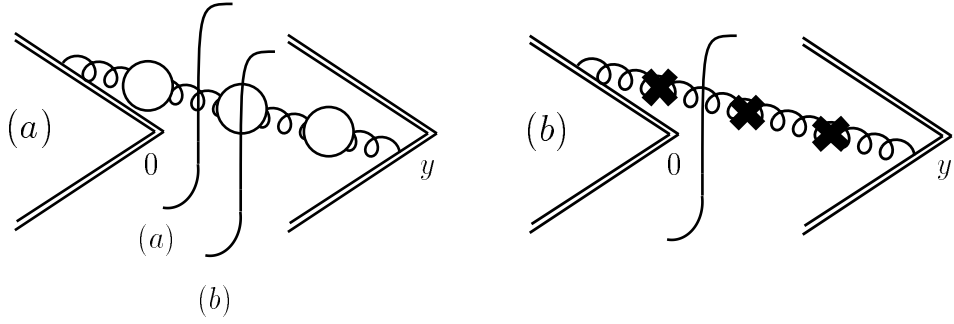}}
   \vspace*{0.3cm}
\caption{\label{wl} (a) Cut diagrams with fermion loops. (b) Non-vanishing
contribution with no cut fermion loop. The crosses denote counterterm
insertions.}
\end{figure}

To compute diagrams with fermion loop insertions into the gluon line,
we follow Appendix~A of \cite{BB94}.
Each fermion loop in a diagram like in Fig.~\ref{wl} is
accompanied by a factor

\be \alpha_s(\mu)\left(-\frac{\beta_0^f}{\eps}\right) L(\eps) \left(
-\frac{k^2}{\mu^2 e^{\gamma_E}}\right)^{-\eps}\,\qquad
L(\eps)\equiv\frac{6\Gamma(1+\eps)\Gamma(2-\eps)^2}{\Gamma(4-2\eps)}
\ee
\phantom{\ref{wl}}

\n and $\beta_0^f=N_f/(6\pi)$. As in lowest order, the only non-vanishing
type of diagram is that shown in Fig.~\ref{wl}a and the symmetric one.
The cuts in this diagram proportional to $\delta(k^2)$ all vanish
except for the diagram with no fermion loop, since with $k^2=0$ the
fermion loop is given by an integral without scale. The diagram with no
loop will be added later. The non-zero
cuts are those proportional to $\Theta(k^2)$, labelled (b), when a
fermion loop is cut. The corresponding imaginary part of the
effective gluon
propagator with an arbitrary number of fermion loops is\footnote{A
pedestrian way of deriving this is to sum and square
the diagrams with one cut
bubble literally. With the correct phases, the sum of all interference
terms in order $\alpha_s^n$ combine as
\begin{displaymath}
 \pi\beta_0^f\frac{6 (1-\eps)\Gamma(2-\eps)}{\Gamma(4-2\eps)}
\sum_{k=0}^{n-1}\cos(n-(2 k+1))\pi\eps = -\frac{\beta_0^f}{\eps}
L(\eps) \sin(-n\pi\eps)\,.
\end{displaymath}}

\be
\frac{g_{\mu\nu}}{k^2}\sum_{n=1}^\infty\alpha_s^n\left(-\frac{\beta_0^f}
{\eps}\right)^n L(\eps)^n \left(\frac{k^2}{\mu^2 e^{\gamma_E}}
\right)^{-n\eps}\sin\pi (-n\eps)\,.
\ee

\n The $k_\mu k_\nu$-terms can be dropped due to gauge invariance
(they are cancelled by diagrams not shown).
A short calculation gives for the sum of all diagrams with cut fermion
loops

\bea
&&\frac{C_F\alpha_s}{\pi}\frac{2}{1-z}\sum_{n=1}^\infty
\alpha_s^n\frac{(\beta_0^f)^n}{(n+1)(-\eps)^{n+1}}\,
G(-\eps,-(n+1)\eps)\,,\\
&&\nonumber G(-\eps,-(n+1)\eps) \equiv L(\eps)^n\left(\frac{Q^2}{4 \mu^2
e^{\gamma_E}} (1-z)^2\right)^{-(n+1)\eps}\frac{\sqrt{\pi}}
{\Gamma(1/2-(n+1)\eps)\Gamma(1+n\eps)}\,.
\eea

\n When we add the counterterms for the fermion loops, a non-vanishing
contribution requires again that at least one fermion loop is cut,
except when {\em all} fermion loops are replaced by counterterms as
in Fig.~\ref{wl}b. The second contribution will be added shortly.
Accounting for the first one, the previous expression is modified to
\cite{BB94}

\be\label{renloops}
\frac{C_F\alpha_s}{\pi}\frac{2}{1-z}\sum_{n=1}^\infty
\alpha_s^n (\beta_0^f)^n\sum_{k=0}^{n-1}\frac{1}{(-\eps)^{n+1}}
\frac{(-1)^k}{n+1-k}\left(\!\!\begin{array}{c}
n\\k\end{array}\!\!\right) G(-\eps,-(n+1-k)\eps)\,.
\ee

\n Next we note that the remaining contribution from the lowest order
diagram together with the ones where all fermion loops are replaced
by counterterms is simply

\be \label{counters}
W^{\rm (1)}_{\rm DY}(z,\alpha_s)\,\frac{1}{1-\alpha_s\beta_0^f/\eps}
\ee

\n with $W^{\rm (1)}_{\rm DY}(z,\alpha_s)$ as in eq.~(\ref{lowestorder}).
Adding this term amounts to extending the sum over $n$ to 0 and
the sum over $k$ to $n$ in eq.~(\ref{renloops}).\footnote{
Let us add an interesting side-remark: For IR safe quantities
like the hard cross section in DY, the contribution from
eq.~(\ref{counters}) is necessary to cancel all IR poles in
$\eps$. However, in this case it is possible to take the
Borel transform of the series, set $\eps$ to zero at this
stage of the calculation and let $u$ act as regulator.
Then eq.~(\ref{counters}) is proportional to
$\exp(-u/\eps)$, which can be set to zero as $\eps\to 0$, provided
we assume $u$ to be (small and) positive. Only the Borel
transform of a term like
eq.~(\ref{renloops}) is left, which apparently does
not contain the lowest order contribution in $\alpha_s$. However, due
to an additional $1/u$, the Borel transform of a term
like eq.~(\ref{renloops})
is in fact of order one close to $u=0$, so that re-expansion of
the Borel integral in $\alpha_s$ produces an order $\alpha_s$
correction, with coefficient equal to the lowest order correction.
It is not difficult to derive eq.~(\ref{rel}) in this way. Alternatively,
one can follow this section, put $\eps=0$ from the very
beginning and use
\begin{displaymath} \int\frac{d^4 k}{(2\pi)^4} =
\intl_0^\infty d\lambda^2
\int\frac{d^3 k}{(2\pi)^3\sqrt{\vec{k}^2+\lambda^2}}\,.
\end{displaymath}
to obtain the Borel transform in the Mellin representation of
eq.~(\ref{rel})}

Eq.~(\ref{renloops}) contains poles $\alpha_s^n/\eps^{n+1}$, but an
additional $1/\eps$ is in fact present as seen by introducing
+-distributions. At this point it is convenient to pass to moments,
before we take the final overall subtractions. The integral over
$z$ is trivial and we obtain

\be\label{wilsonmom}
W_{\rm DY}(N,\alpha_s) = 2 \frac{C_F\alpha_s}{\pi}
\sum_{n=0}^\infty
\alpha_s^n (\beta_0^f)^n\sum_{k=0}^{n}\frac{1}{(-\eps)^{n+2}}
\frac{(-1)^k}{(n+1-k)^2}\left(\!\!\begin{array}{c}
n\\k\end{array}\!\!\right) H(-\eps,-(n+1-k)\eps)\,,
\ee
\be
H(\eps,s)= \left[L(-\eps)\right]^{s/\eps-1}\left(\frac{Q^2}{\mu^2
e^{\gamma_E}}\right)^{s}\frac{\Gamma(N)\Gamma(1+s)}
{2\Gamma(N+2 s)\Gamma(1-s+\eps)}\,.
\ee

\n Note the signs of the arguments of the function $H$. We introduce
expansions

\be\label{Hexpansion}
H(\eps,s) = \sum_{j=0}^\infty H_j(\eps) \,s^j\,,\qquad
H_j(\eps) = \sum_{k=0}^\infty h_k^{[j]} \eps^k\,.
\ee

\n The functions $H_j$ are finite for $\eps\to 0$ by construction. For later
use we collect the expressions

\bea\label{collect}
H(0,s) &=& \left(\frac{Q^2}{\mu^2} e^{-5/3}\right)^s \frac{\Gamma(N)
\Gamma(1+s)}{2\Gamma(N+2 s)\Gamma(1-s)}\nonumber\\
H_0(\eps) &=& \frac{1}{2}\frac{\Gamma(4+2\eps)}{6\Gamma(1-\eps)
[\Gamma(2+\eps)]^2\Gamma(1+\eps)}\\
H_1(\eps) &=&\nonumber H_0(\eps)\left\{\frac{1}{\eps}\ln L(-\eps)
+\ln\frac{Q^2}{\mu^2 e^{2\gamma_E}}-2\Psi(N)+\Psi(1+\eps)\right\}\,.
\eea

\n To subtract the overall divergences from eq.~(\ref{wilsonmom}),
$H(-\eps,-(n+1-k)\eps)$ is first expanded in its second variable

\bea
W_{\rm DY}(N,\alpha_s) &=& 2 \frac{C_F\alpha_s}{\pi}
\sum_{n=0}^\infty
\alpha_s^n (\beta_0^f)^n\sum_{j=0}^\infty\frac{H_j(-\eps)}
{(-\eps)^{n+2-j}}
\sum_{k=0}^{n}(-1)^k\left(\!\!\begin{array}{c}
n\\k\end{array}\!\!\right) (n+1-k)^{j-2}\nonumber\\
&=& 2 \frac{C_F\alpha_s}{\pi} \sum_{n=0}^\infty
\alpha_s^n (\beta_0^f)^n \Bigg\{\frac{(-1)^n}{n+1} (\Psi(n+2)-\Psi(1))
\frac{1}{(-\eps)^{n+1}} H_0(-\eps)\\
&&+\,\frac{(-1)^n}{n+1}\frac{1}{(-\eps)^{n+1}} H_1(-\eps)+n!\,H_{n+2}(0)
+ {\cal O}(\eps)\Bigg\}\,.\nonumber
\eea

\n To arrive at the second line we use

\be
\sum_{k=0}^{n}(-1)^k\left(\!\!\begin{array}{c}
n\\k\end{array}\!\!\right) (n+1-k)^{j-2} =
\left\{\,\,\,\begin{array}{ll}
(-1)^n/(n+1) (\Psi(n+2)-\Psi(1)) & j=0\\
(-1)^n/(n+1) & j=1\\
0 & j=2,\ldots,n+1\\
n! & j=n+2\end{array}\right.
\ee

\n Minimal subtraction of the pole part in $\eps$ gives the final expression
for the $\overline{\rm MS}$-renormalized Wilson line with an infinite
number of fermion loops:

\bea\label{wilsren}
W_{\rm DY}(N,\alpha_s) &=& 2 \frac{C_F\alpha_s}{\pi}
\sum_{n=0}^\infty
\alpha_s^n (\beta_0^f)^n \bigg\{\frac{(-1)^n}{n+1} (\Psi(n+2)-\Psi(1))
\,h^{[0]}_{n+2}\nonumber\\
&&\hspace*{2cm}+\,\frac{(-1)^n}{n+1}\,h^{[1]}_{n+1}+n!\,H_{n+2}(0)
\bigg\}
\eea

\n Note that the result
contains two terms related to the `anomalous dimension functions'
$H_0$ and $H_1$, that appeared as coefficients of the $1/\eps^{n+2}$
and $1/\eps^{n+1}$ pieces before, and a `finite term' with coefficient
$n!$. It is instructive to take the Borel transform of the above
series. We then obtain

\be\label{Wloop2}
B\left[W_{\rm DY}\right](N,u) = 2 \frac{C_F}{\pi}\left[
\frac{H(0,-u)-H_0(0)+u \tilde{H}_1(u)}{u^2} + \frac{1}{u}\frac{d}{d u}
\intl_0^1 d x\frac{x \tilde{H}_0(u)-\tilde{H}_0(x u)}{x (1-x)}\right]\,,
\ee

\n where $\tilde{H}_j(u)\equiv\sum_{n=0} h_n^{[j]} u^n/n!$ is essentially
the Borel transform of $H_j(\epsilon)$, cf. eq.~(\ref{Hexpansion}).
With the explicit expression for $H(0,-u)$ from eq.~(\ref{collect}),
we find that this term above coincides with the large-$N$
limit of the Borel transform of the full Drell-Yan cross section in
eq.~(\ref{leadingn}), up to corrections $1/N$ that have been
neglected in both cases. There it was understood that the double pole at
$u=0$ would be cancelled by virtual corrections and distribution
functions, but eq.~(\ref{leadingn}) provided the leading power corrections
$(N\lqcd/Q)^{2 k}$ ($k=1,\ldots$). In the present case the divergences
are subtracted by the functions $H_0$ and $H_1$. These additional terms
are analytic in $u$ (as expected for anomalous dimensions in minimal
subtraction schemes), so that indeed the Wilson line operators
incorporate the leading power corrections (with the largest positive
power of $N$) to the hard Drell-Yan cross section. In particular,
no power correction of order $1/Q$ appears. Another important
observation is that only the function $H_0$ is related to the most
singular pole in $\eps$, which comes from soft and collinear
emission. While $H_0$ will therefore be related to the universal
cusp (eikonal) anomalous dimension, all conclusions on power corrections
follow from $H(0,-u)$ which is process-specific (that is, depends
on the Wilson contour and not its cusps alone).

\subsection{Evolution equation and resummation}

Since eq.~(\ref{wilsren}) is exact in the single-chain
approximation, it must satisfy the RG equation (\ref{evolutionequation})
to the same accuracy, that is with
$\beta(\alpha_s)=\beta_0\alpha_s^2$ and $\ln [1+W_{\rm DY}]$
equivalent to $W_{\rm DY}$,
 see the discussion in Sect.~2.3. (The unity under the
logarithm is
the tree-level contribution $W_{\rm DY}^{(0)}=1$.)
Using the explicit expression in eq.~(\ref{Wloop2}) we get

\bea\label{rhs1}
&&\left(\mu^2\frac{\partial}{\partial\mu^2} + \beta_0\alpha_s^2
\frac{\partial}{\partial \alpha_s}\right)\,\ln W_{\rm DY}
=
-\frac{1}{\beta_0}\int_0^\infty du\, e^{u/(\beta_0 \alpha_s)}
\left[\mu^2\frac{\partial}{\partial\mu^2} -u\right]
B[W_{\rm DY}](N,u)
\nonumber\\
&&\hspace*{2cm}=
\frac{2 C_F \alpha_s}{\pi}\left\{-H_1(a_s)-\sum_{n=0}^\infty a_s^n
h^{[0]}_{n+1}[\Psi(n+2)-\Psi(1)]\right\}
\nonumber\\
&&\hspace*{2cm}=
\frac{2 C_F \alpha_s}{\pi}\left\{-H_1(a_s)-\frac{1}{a_s}
\int_0^1 dx\, \frac{H_0(a_s)-H_0(x a_s)}{1-x}\right\}\,,
\eea

\n where we have restored the full non-abelian
$\beta_0=-(11-2/3 N_f)/(4\pi)$
and introduced

\be
a_s =-\beta_0 \alpha_s\,.
\ee

\n In the large-$N$ limit we expect to find a term logarithmic in $N$,
multiplied by the cusp anomalous dimension. From the explicit expression
for $H_1$ we get indeed

\be
-\frac{2 C_F \alpha_s}{\pi}H_1(a_s) =
\frac{C_F \alpha_s}{\pi}
\frac{\Gamma(4+2a_s)}{6 \Gamma(1-a_s)[\Gamma(2+a_s)]^2\Gamma(1+a_s)}
\cdot 2 \ln\frac{N\mu}{N_0 Q}
\ee

\n with $N_0=e^{-\gamma_E}$ and omitting terms without $\ln N$. Thus

\bea\label{GammaCusp}
\Gamma_{\rm cusp}(\alpha_s) &=&
2\frac{C_F \alpha_s}{\pi} H_0(a_s) = \frac{C_F \alpha_s}{\pi}
\frac{\Gamma(4+2a_s)}{6 \Gamma(1-a_s)[\Gamma(2+a_s)]^2\Gamma(1+a_s)}
\nonumber\\
&=&\frac{C_F \alpha_s}{\pi}\left\{
1-\frac{5}{3}a_s-\frac{1}{3} a_s^2+2.0708 a_s^3-1.0932 a_s^4 +\ldots
\right\}\,,
\eea

\n in agreement with  the known two-loop result (keeping the leading
$N_f$-term only) \cite{KOD82,KOR87}. We verify coincidence with
the direct calculation in Appendix A.

The remaining terms in eq.~(\ref{rhs1}) (up to $1/N$-corrections,
which are consistently neglected)
give $\Gamma_{\rm DY}(\alpha_s)$:

\bea\label{GammaDY}
  \Gamma_{\rm DY}(\alpha_s) &=& 2\frac{C_F \alpha_s}{\pi}
\left\{H_0(a_s)\left[-\frac{1}{a_s}\ln L(-a_s)-\Psi(1+a_s)\right]
-\sum_{n=0}^\infty a_s^n h^{[0]}_{n+1}[\Psi(n+2)-\Psi(1)]\right\}
\nonumber\\
&=&2 \frac{C_F \alpha_s}{\pi}\left\{
\left(\frac{7}{9}-\frac{\pi^2}{24}\right)a_s
-0.08547 a_s^2-0.1537 a_s^3+0.4951 a_s^4 +\ldots\right\}
\eea

\n Note that $\Gamma_{\rm DY}(\alpha_s)$ starts from second order in
$\alpha_s$ \cite{KOR93}. Eq.~(\ref{Wloop2}) can now be written as

\be
B[W_{\rm DY}](N,u) = B[W^{\rm real}](N,u)\left(\frac{\mu^2}{Q^2}\right)^u
-\frac{1}{u^2} B[\Gamma_{\rm cusp}](u)\left[1+2 u\ln\frac{N\mu}{N_0 Q}
\right]-\frac{1}{u} B[\Gamma_{\rm DY}](u)
\ee

\n The first term on the right hand side gives the
unrenormalized Borel transform, which coincides with $B[W^{\rm real}]$,
eq.~(\ref{leadingn}), and the two other terms are the subtractions,
necessary for finiteness at $u=0$.

The anomalous dimensions, eq.~(\ref{GammaCusp}) and (\ref{GammaDY}),
have finite radius of convergence (in the $\overline{\rm MS}$ scheme)
and their Borel transforms are entire functions. At $u=1/2$ they
are non-zero.
We conclude that none of the terms in the evolution equation for the
Wilson line, eq.~(\ref{evolutionequation}), contains renormalons,
and neither does the particular solution corresponding to the
integral in eq.~(\ref{sol1}). (Let us insist again that one must stay
in the perturbative domain $N<Q/\lqcd$).
In the present formalism, all IR renormalons to resummed
cross sections are introduced through the initial condition (or solution
to the homogeneous equation) for
the evolution of the Wilson line. The initial condition can be determined
by comparison with exact calculation (in the single-chain approximation),

\be \label{ndep}
B[X](N,u) = B[W^{\rm real}](N,u)-\frac{1}{u^2}\left(\frac{N}{N_0}
\right)^{2 u} \left[B[\Gamma_{\rm cusp}](u)+u B[\Gamma_{\rm DY}](u)
\right]\,,
\ee

\n and consequently, the renormalons in the initial condition are
exactly those in eq.~(\ref{Wloop2}). We find once more that
the dominant IR renormalon requires non-perturbative contributiuons only
at order $N^2\Lambda_{\rm QCD}^2/Q^2$.
All terms in eq.~(\ref{ndep}) have the same $N$-dependence proportional
to $N^{2 u}$. This verifies that the initial condition is a function
only of $\alpha_s(Q N_0/N)$. Indeed, using eq.~(\ref{homogen}) we get

\be
B[X(\alpha_s(QN_0/N)](u) = \left(\frac{N}{N_0}\right)^{2u}
B[X(\alq)](u)\,.
\ee

It remains to clarify the origin of apparent linear power corrections
to the resummed cross section in the form of eq.~(\ref{exponent}).
The difference with eq.~(\ref{sol1}) lies in the choice of
the inhomogeneous term in the solution to the evolution equation
as

\bea \label{sol2}
\ln W_{\rm DY}\!
\left(\frac{\mu^2 N^2}{Q^2 N_0^2},\alpha_s(\mu)\right) &=&
X'(\alpha_s(Q N_0/N))\,
-2\intl_0^{1} d z\frac{z^{N-1}-1}{1-z}
\Bigg[\intl_{(1-z)^2 Q^2}^{\mu^2}
\frac{d k_t^2}{k_t^2}\,\Gamma_{\rm cusp}(\alpha_s(k_t)
\nonumber\\
&&\hspace*{3cm}
+\,\Gamma_{\rm DY}(\alpha_s((1-z) Q)\Bigg]\,.
\eea

\n A simple calculation (as in Sect.~2.2) yields for the
difference of Borel transforms of the integrals in eq.~(\ref{sol1}) and
eq.~(\ref{sol2})

\be
\left(\frac{N}{N_0}\right)^{2 u} \left(N_0^{-2 u}\Gamma(1-2 u)
-1\right)\frac{1}{2 u^2}\left[B[\Gamma_{\rm cusp}](u)+u B[
\Gamma_{\rm DY}](u)\right]\,.
\ee

\n This difference has to be compensated by
a change in the solution of the homogeneous RG equation, which
now becomes,

\be
B[X'](N,u) = B[W^{\rm real}](N,u)-\frac{1}{u^2}\left(\frac{N}{N_0}
\right)^{2 u} N_0^{-2 u}\Gamma(1-2 u)
\left[B[\Gamma_{\rm cusp}](u)+u B[\Gamma_{\rm DY}](u)
\right]\,.
\ee

\n Comparing with eq.~(\ref{ndep}), we see that this change involves
singular terms at $u=1/2$.
Because of the $\Gamma(1-2 u)$, both, the Borel transform
of $X'$ and of the integral in eq.~(\ref{sol2}) have IR renormalon poles
at $u=1/2$
(corresponding to $1/Q$) with residue proportional to $N$, but opposite
sign, so that in the sum the singularity cancels.
 In the
 Borel transform of the integral in
eq.~(\ref{sol2}) the pole is present through the functions $\Delta_1$
and $\Delta_3$, introduced in Sect.~2.2, and is cancelled
in this representation by an explicit pole in $X'$. If, to return to
the standard representation,
the initial condition contribution $X'$ is eliminated
by the redefinitions of eq.~(\ref{redef}), this cancellation persists
in the exponent. Indeed, for the functions $A$ and $C$ in
eq.~(\ref{exponent}) we get

\bea
B[A](u) &=& B[\Gamma_{\rm cusp}](u)\nonumber\\
-\frac{1}{2} B[C](u) &=& B[\Gamma_{\rm DY}](u) + u \left(\frac{N_0}{N}
\right)^{2 u}\frac{N_0^{2 u}}{\Gamma(1-2 u)} B[X](N,u)\\
&=& u B[W^{\rm real}](N,u)\left(\frac{N_0}{N}
\right)^{2 u}\frac{N_0^{2 u}}{\Gamma(1-2 u)} -\frac{1}{u} B[
\Gamma_{\rm cusp}](u)\nonumber\,,
\eea

\n so that

\be 2 B[A](u) - u B[C](u) = \frac{2 N_0^{2 u}}{\Gamma(1-2 u)} u^2
\left(\frac{N_0}{N}\right)^{2 u} B[W^{\rm real}](N,u)\,.
\ee

\n Due to the Gamma-function in the denominator, this combination has
a zero at $u=1/2$, independent of the particular form of
$B[W^{\rm real}](N,u)$ (unless it is singular, which is not the
case, at least in the single-chain approximation, see eq.~(\ref{Wloop2})).
Note that all contributions from the anomalous dimensions
$\Gamma_{\rm cusp}$
and $\Gamma_{\rm DY}$ disappear.

We conclude that the cancellation of scenario B in Sect.~2.2 does
occur and no indication for physical `higher twist' corrections
of order $1/Q$ remains. The cancellation of such corrections with the
remainder $R$ as in mechanism C could only have taken place, if the
soft approximation had accounted correctly for the leading-$N$
behaviour only to logarithmic, but not to power-like accuracy. This
is not the case. As noted before (and expected), $W_{\rm DY}$
computed in the eikonal approximation (Wilson line) reproduces those
power corrections to the Drell-Yan hard cross section with the
largest power in $N$.
This suggests that the nonperturbative corrections indicated by
renormalons in the Wilson line should exponentiate \cite{KOR95}.

%%%%%%%%%%%%%%%%%%%%%%%%%%% SECTION 6 %%%%%%%%%%%%%%%%%%%%%%%%%%%%%%%%%

\section{Discussion and Summary}
\setcounter{equation}{0}

Let us summarize our investigation of power corrections (alias infrared
renormalons) in Drell-Yan production. Our starting observation is that
the moments of the resummed hard cross section in the large-$N$ limit
can be written in two different ways as

\bea\label{form11}
\ln \omega_{q\bar{q}}(N,\alq) &=& -\intl_0^1 d z\,\frac{z^{N-1}-1}
{1-z} \Bigg\{2\intl_{Q^2 (1-z)}^{Q^2 (1-z)^2} \frac{d k_t^2}{k_t^2}\,
\Gamma_{\rm cusp}(\alpha_s(k_t)) + B(\alpha_s(\sqrt{1-z} Q))\nonumber\\
&& \hspace*{2cm} +\,C(\alpha_s((1-z) Q)\Bigg\} + {\cal O}(1)
\eea

\n or

\bea\label{form2}
\ln \omega_{q\bar{q}}(N,\alq) &=& \intl_0^{1-e^{-\gamma_E}/N}
d z\,\frac{1}{1-z} \Bigg\{2\intl_{Q^2 (1-z)}^{Q^2 (1-z)^2}
\frac{d k_t^2}{k_t^2}\,\Gamma_{\rm cusp}(\alpha_s(k_t)) +
\tilde{B}(\alpha_s(\sqrt{1-z} Q))\nonumber\\
&& \hspace*{2cm} +\,\tilde{C}(\alpha_s((1-z) Q)\Bigg\} + {\cal O}(1)\,.
\eea

\n Both forms are equivalent for the purpose of summing large logarithms
in $N$, but have apparently different properties as far as sensitivity
to the IR Landau pole of the coupling is concerned. When the `anomalous
dimensions' $\Gamma_{\rm cusp}$, $B$ etc. are truncated at finite order,
the second form has no ambiguity unless $N>Q/\lqcd$, whereas such an
ambiguity is present in the first version for any $N$. This apparent
conflict is resolved by the analysis of the exponent in Sect.~2.2, which
concludes that the correct IR renormalon structure of the exponent can
be obtained only, if the anomalous dimensions are also considered to
all orders. With respect to eqs.~(\ref{form11}), (\ref{form2}) this means
that IR renormalons that are manifest in terms of the integrals over
$z$ and $k_t$ (the functions $\Delta_i$ in Sect.~2.2)
in one form can be hidden in the large-order behaviour
of the anomalous dimension functions in the other. It is with this
observation that we start to differ from previous works,
which implicitly assumed that low-order approximations
to the anomalous dimensions would be sufficient \cite{CON94a} or
the functions $B$ and $C$, which account for gluon emission that is
not both, soft and collinear, could be neglected \cite{KOR95,Ktalk}.

Already from this conclusion it is evident that, even in the large-$N$
limit the search for power corrections is a problem quite different from
resummation of $\ln N$, which can be completed up to a desired order
with finite-order approximations to the anomalous dimensions. Although
the techniques of soft factorization developed for the purpose of
resummation are still effective in isolating those power corrections
$(N\lqcd/Q)^{2 k}$ with the largest power of $N$, this task requires
application of these techniques to all orders in perturbation theory.

We have argued that taking into account these considerations, all
corrections of order $1/Q$ cancel in the Drell-Yan cross section
and the remaining IR renormalons indicate parametrically smaller
power corrections of order $N^2\lqcd^2/Q^2$. The evidence for this
cancellation has been collected from two sources: A calculation of the
Drell-Yan cross section for any $N$ with finite gluon mass as IR
regulator (which by eq.~(\ref{rel}) can be considered also as a
certain approximation to large orders in perturbation theory) does not
reveal any IR sensitivity proportional to $\lambda/Q$, even though
the phase space in the presence of finite gluon mass is reduced
linearly in $\lambda/Q$. The same result has been obtained from
implementing the factorization of soft gluon emission in terms of
Wilson lines \cite{KOR93} in the approximation of a single chain
of vacuum polarization insertions. This approximation may be
considered as a model (exact in the unphysical limit of large number
of fermions)
for the large-order behaviour of the anomalous dimension functions.
In this language the absence of $1/Q$-power corrections seems to be
most transparent. Renormalizing the Wilson line by minimal subtraction,
we find that the anomalous dimensions that enter the renormalization
group equation for the Wilson line are analytic for small $\alpha_s$
and do not contain IR renormalons. These enter the solution of the
evolution equation exclusively through the initial (boundary)
condition, which,
in general, contains both, perturbative and non-perturbative contributions.
As in the exact calculation of the hard Drell-Yan cross section, the
power corrections deduced from IR renormalons are of order
$N^2 \lqcd^2/Q^2$. It is only when this initial condition is absorbed
into a redefinition of the functions $B$ and $C$ that IR renormalons
enter the exponent. These redefintions are such that the power
corrections of order $1/Q$ found from eq.~(\ref{form11})
\cite{CON94a,KOR95} are eliminated by an exact zero in the Borel
transforms of the corresponding anomalous dimensions.

In Sect.~4 we have seen that the identification of power corrections
through IR renormalons or cutoff dependence relies on an exact treatment
of soft gluon emission also at large angles, while the collinear
approximation can introduce spurious cutoff dependence. It is for
this reason that the leading (in $N$) power corrections can not be
described by the universal cusp anomalous dimension. In terms of
Wilson lines this is seen explicitly by the fact that IR renormalons
enter through the initial condition and not as the coefficients
of the leading soft and collinear pole in $\eps$. We also note that
the result on power corrections is not specific to the semi-inclusive
limit and applies to all $N$ in the perturbative domain
$N<Q/\lqcd$. The analysis of Wilson lines suggests however that, similarly
to logarithms in $N$, the power corrections leading in $N$ can
also exponentiate.

All explicit calculations of large-order behaviour have been done
in the approximation of a single `renormalon chain'. If we consider
this approximation as an approximation for $\ln \omega_{q\bar{q}}$
rather than the hard cross scetion $\omega_{q\bar{q}}$ as suggested
by exponentiation of large logarithms beyond this approximation, then
as usual the contributions from several chains are bound to modify
the precise numerical coefficient of the $N^2\lqcd^2/Q^2$-terms.
To confirm or disprove the absence of $1/Q$ contributions to the
Drell-Yan process at the level of two gluon emission would again
require an exact treatment of soft gluon emission that goes beyond
the usual approximations in treating parton cascades. From the fact
that even in the single-chain approximation the IR renormalon
structure is sensitive to the large-order behaviour of the
anomalous dimensions, we expect this to be a quite challanging
problem. Without any further physical evidence to the opposite
we find it reasonable to
assume that as in other cases (with operator product expansion
as a check) \cite{irr} multiple chains and more complex diagrams
will not modify the power behaviour itself. Still, we recall that
the Drell-Yan process may receive power corrections, which can not
be seen through IR renormalons, and whose size is not accessible
with present theoretical tools.

As pointed out in \cite{DOK95,AKH95},
the question of linear power corrections is of particular importance
due to their potential universality, which would allow to treat
such power corrections to processes without operator product expansion
on a phenomenological level akin to matrix elements of higher dimensional
operators in cases with operator product expansion.\footnote{The statement
of universality implies that the universality of IR renormalons is
also preserved non-perturbatively. This is not always the case. For
example, the `binding energy' $\bar{\Lambda}$ that appears in
heavy quark expansions has a universal IR renormalon ambiguity, but
its numerical value (after proper definition) depends on the particular
hadron under consideration. Similarly, one would expect the actual
value of corrections of order $1/Q^2$ in the Drell-Yan process to
depend on the particular hadron target.} From this point of view it
is interesting to compare the result on power corrections for the
Drell-Yan process with those for hadronic event shape variables
in $e^+ e^-$-annihilation \cite{MWI95,WEB94,DOK95,AKH95}. In particular,
from the point of view of summing large logarithms in $N$ to next-to-leading
accuracy, the thrust distribution is closely related to the distribution
in $z$ in the Drell-Yan process, $\omega_{q\bar{q}}(z,\alq)$, since the
large logarithms can be accounted for by the same jet mass
distribution \cite{CAT91b,CAT93}. (For a discussion in the present
context, see \cite{AKH95,Ktalk}.) On the other hand, if we compute
the thrust distribution with finite gluon mass to order $\alpha_s$,
we obtain for the averages

\bea
\langle 1-T\rangle &\sim& -8 G\,\frac{C_F\alpha_s}{\pi}\frac{\lambda}
{Q}\nonumber\\
\langle T^{N-1}\rangle &\sim& 8 G\,(N-1)\,\frac{C_F\alpha_s}{\pi}
\frac{\lambda}{Q}\,,
\eea

\n where $G=\sum_{k=0} (-1)^k/(2 k+1)^2=0.91596\ldots$ is Catalan's
constant.\footnote{The discrepancy between the coefficient $-8 G$ and
the value $-4$ reported in \cite{WEB94} arises entirely from the
normalization of thrust. In the region where $x_1+x_2$ ($x_1$, $x_2$
being the
energy fractions of quark and antiquark) is close to its maximum
value, we have, for finite $\xi=\lambda^2/Q^2$,
\begin{displaymath}
T=\max_{\vec{n}}\frac{\sum_i |\vec{p}_i \vec{n}|}{\sum_i |\vec{p}_i|}
= \frac{2 \max(x_1,x_2)}{x_1+x_2+\sqrt{(2-x_1-x_2)^2-4\xi}}\,.
\end{displaymath}
One can not set $\xi=0$ in the square root, since in the region
of interest $(2-x_2-x_2)^2$ is of order $\xi$. One can see that
the difference is a $\sqrt{\xi}$-effect, since with the above
normalization $1-T=0$ in the two-jet region $x_1+x_2=2 (1-\sqrt{\xi})$,
while with $\xi=0$ in the expression for $T$, one would obtain
$1-T=\sqrt{\xi}$. See also Appendix~B.} Contrary to the Drell-Yan
process, large $1/Q$-corrrections are present for event shapes,
in particular for the moments of the thrust-distribution. Some caution
is necessary in interpreting the results obtained with finite
gluon mass in these cases, since the correspondence with IR renormalons,
provided by eq.~(\ref{rel}) for the DY process, does not hold, because
the phase space of the quark pair from gluon splitting in the
single chain approximation is not integrated unweighted. We refer to
the very recent publication \cite{NAS95}, where the relevance of
four-parton contributions to $1/Q$-terms is discussed. The difference
in $1/Q$-corrections to the DY cross section and thrust reinforces
that these corrections can only be identified by a treatment of soft
gluon emission at all angles to all orders.

One may speculate on a deeper physical reason for this difference.
One possibility is that IR safety of event shape variables is in a
certain sense enforced `by hand'. Depending on the construction,
this can enhance or suppress IR regions in an almost arbitrary
way \cite{MWI95}. On the other hand, for the DY process one might be
more confident that IR renormalons probe some aspect of the IR dynamics
of QCD that eventually might find a rigorous treatment just as
higher twist matrix elements in deep inelastic scattering. In this
respect it is somewhat comforting that corrections to the DY process
appear to be of order $1/Q^2$.

The theoretical discussion has been performed in terms of moments.
To obtain a prediction for the resummed DY cross section, the inverse
Mellin transform has to be taken. Let us note first that there is
in fact no compelling reason to evaluate the integrals with running
coupling in eqs.~(\ref{form11}) and (\ref{form2}) exactly as long as,
as always in practice, only finite-order approximations to the
anomalous dimensions are available. To any definite accuracy in the
summation of logarithms of $N$, one may expand $\alpha_s(k_t)$ etc.
to the required order
and perform the integrals term by term. Then the Landau pole in the
perturbative running coupling never poses a difficulty. This reflects
our previous assertion that the question of power corrections
necessarily deals with the all-order behaviour of the anomalous
dimensions.

Nevertheless the exact integrals over the running coupling are often
kept and the dependence on regulating the Landau pole is then interpreted
as the ultimate accuracy that can be achieved without deeper understanding
of non-perturbative corrections. From our analysis we conclude that such
a treatment has to be interpreted with care, since the failure to
include the anomalous dimensions to all orders may pretend artificially
large regularization dependence. From this point of view, the
construction of the inverse Mellin transform starting from
eq.~(\ref{form2}) is preferred compared to eq.~(\ref{form11}), which was
used in \cite{CON94a}, since spurious large corrections of order
$N\lqcd/Q$ do not appear. We find confirmation for this suggestion
from the analysis of \cite{ALV94}, where comparison of the resummed
cross section based on eq.~(\ref{form11}) \cite{CON94a,ALV95} with
data requires to fit rather large corrections proportional
to $N\lqcd/Q$, whose main effect is to reduce the effect of resummation.

The phenomenological implications could pertain to processes other
than Drell-Yan. For instance,
a resummation similar to the Drell-Yan case has also been completed
for the top production cross section (where in addition large corrections
from gluon-gluon fusion have to be considered) \cite{LAE92}. It might
be worthwhile to investigate the rather large cutoff dependence reported
in the light of the results presented in this paper, especially
since the cutoff was chosen much larger than $\lqcd$. An improvement
of the theoretical Standard Model prediction is crucial to unravel
potential anomalous couplings of the top quark
in its production cross section.

We conclude that further explorations of power corrections, both theoretical
and phenomenological, are expected to provide further insight into the
workings of QCD and will help understanding the limitations of
perturbative QCD in describing data for hadronic observables.

\vspace*{1cm}
{\bf Acknowledgements.} M.~B. thanks G.~Marchesini and I.~Z.~Rothstein
and V.~B. thanks G.~P.~Korchemsky for helpful discussions and critical
reading of the manuscript. Both authors
are grateful to the members of the CERN Theory group for their hospitality,
when part of this work was done. M.~B. is supported by the
Alexander von Humboldt-foundation.

%%%%%%%%%%% APPENDIX A
\setcounter{equation}{0}
\setcounter{section}{0}
\renewcommand{\theequation}{A.\arabic{equation}}

\vspace*{0.7cm}
\noindent {\Large\bf Appendix A}\\

\begin{figure}[t]
   \vspace{0cm}
   \centerline{\epsffile{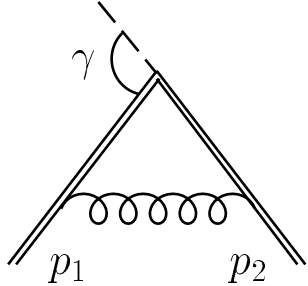}}
   \vspace*{0.3cm}
\caption{\label{cusp} Diagram for the cusp anomalous dimension. The
double lines denote a Wilson line with cusp (alternatively, they
can be interpreted as eikonal propagators).}
\end{figure}

\noindent In this appendix we sketch a simple derivation of
the cusp anomalous
dimension to all orders in the $1/N_f$-approximation (single-chain
approximation). In lowest order, the cusp anomalous dimension is
given by the $1/\eps$-pole in the diagram shown in Fig.~\ref{cusp}.
In higher orders we adopt dimensional regularization with minimal
subtraction. As shown in \cite{BBB1} the series in the
$1/N_f$-approximation can be extracted from the large-$\lambda$
behaviour of the diagram, when evaluated with a gluon of mass
$\lambda$. A straightforward calculation gives

\be
-i g_s^2 C_F\mu^{2\eps} \int\frac{d^d k}{(2\pi)^d}
\frac{p_1\cdot p_2}{(k^2-\lambda^2) \,p_1\cdot k\,p_2\cdot k} =
-\frac{C_F\alpha_s}{2\pi}\frac{\Gamma(1+\eps)}{\eps}\left(
\frac{\lambda^2}{4\pi\mu^2}\right)^{-\eps}\,\gamma\,\mbox{coth}\,
\gamma\,,
\ee

\n where $\mbox{cosh}\,\gamma=(p_1\cdot p_2)/\sqrt{p_1^2 p_2^2}$ is
the Minkowskian cusp angle \cite{KOR87}. The cusp anomalous dimension
can be read off the coefficient of $1/\eps\,(\lambda^2/\mu^2)^{-\eps}$
\cite{BBB1} (Appendix~A) and we obtain (adding self-energy contributions,
which result in $\gamma\,\mbox{coth}\,
\gamma\to \gamma\,\mbox{coth}\,
\gamma-1$)

\bea
\Gamma_{\rm cusp}(\alpha_s) &=& \frac{C_F\alpha_s}{\pi}\,
\frac{\Gamma(4-2\beta_0\alpha_s)}{6\Gamma(2-\beta_0\alpha_s)^2
\Gamma(1-\beta_0\alpha_s)\Gamma(1+\beta_0\alpha_s)}\,
(\gamma\,\mbox{coth}\,\gamma-1)\nonumber\\
&=&\left[C_F\frac{\alpha_s}{\pi}-\frac{5}{18} C_F N_f\left(
\frac{\alpha_s}{\pi}\right)^2-\frac{1}{108} C_F N_f^2\left(
\frac{\alpha_s}{\pi}\right)^3+\ldots\right]
(\gamma\,\mbox{coth}\,\gamma-1)\,,
\eea

\n where $\beta_0=N_f/(6 \pi)$. The second order coefficient is
in agreement with the $N_f$-term of the complete two-loop
result \cite{KOD82,KOR87}. The above series gives the term with
the largest power of $N_f$ to all orders.\\[0.5cm]

\phantom{\ref{cusp}}

%%%%%%%%%%%% APP.B

\setcounter{equation}{0}
\renewcommand{\theequation}{B.\arabic{equation}}

\noindent {\Large\bf Appendix B}\\

\noindent The different power corrections to the Drell-Yan cross
section and
averages of thrust can be reduced to a few elementary integrals. First
we write for the moments

\be \langle X^{N-1}\rangle = \langle 1 \rangle - (N-1) \langle 1-X\rangle
+ \ldots\,,
\ee

\n where $X=T$ (thrust) or $X=z$ (Drell-Yan)
and the brackets denote averaging with the
corresponding distribution in $T$ or $z$. For the Drell-Yan cross section
it is given by eq.~(\ref{realDY}). The omitted terms do not contribute
to terms of order $\lambda/Q$, since each power of $1-X$ suppresses the
region of phase space $X\to 1$ where linear terms can arise. One can show
that $\langle 1\rangle$ is free from such terms in both cases, so that the
problem reduces to the average of $1-X$. This average receives no virtual
corrections and is free from soft divergences.

For the Drell-Yan cross section, we use the new variable $y=1-z-\xi z$
($\xi=\lambda^2/Q^2$). Neglecting all terms that can not give rise to
$\sqrt{\xi}$ in the expansion for small $\xi$, we find

\be \langle 1-z\rangle\sim 4\frac{C_F\alpha_s}{\pi}\intl_{\sqrt{\xi}}^\eps
d y\,\ln\frac{y+\sqrt{y^2-\xi}}{y-\sqrt{y^2-\xi}}
\sim 4\frac{C_F\alpha_s}{\pi}\left[\eps
\ln\xi + 0\cdot\sqrt{\xi}+\ldots\right]\,,
\ee

\n where $\eps$ is an arbitrary cutoff that does not affect linear
terms in $\lambda$ and `$\sim$' means that the expression is accurate
to linear terms. Although the lower limit of the integral is $\sqrt{\xi}$,
its expansion does not contain such terms. The logarithm of $\xi$ is
due to the fact that the partonic cross section is not collinear safe.
It is subtracted by the parton distribution function, which does not
introduce linear terms either (Sect.~3).

In case of thrust it is convenient to switch from
the energy fractions $x_1$ and $x_2$ of quark and anti-quark as integration
variables to $y=1-1/2\,(x_1+x_2)$ and $x=(x_1-x_2)/2$. We then find after
integration over $x$

\bea
\langle 1-T\rangle &\sim& 2\frac{C_F\alpha_s}{\pi}\intl_{\sqrt{\xi}}^\eps
d y\,\left[\left(1-\frac{\xi}{y^2}\right)^{1/2}
\ln\frac{y+\sqrt{y^2-\xi}}{y-\sqrt{y^2-\xi}} +\ln\frac{\xi}{y^2}\right]
\nonumber\\
&\sim& 2\frac{C_F\alpha_s}{\pi}\left[0\cdot
\ln\xi - 4G\sqrt{\xi}+\ldots\right]\,,
\eea

\n where $G=0.91596\ldots$ is Catalan's constant. The additional logarithm
in the integrand as compared to the Drell-Yan case
eliminates the logarithm of $\xi$ as required by
collinear safety of thrust, but introduces a (n additional)
linear correction in
$\lambda/Q$. If the $\xi$-dependence of the normalization $\sum_i |\vec{p}
_i|$ of thrust
is neglected, the square root in front of the logarithm is absent and
we obtain the different result of \cite{WEB94}.

\end{document}